\documentclass[10pt,twocolumn]{article} 
\usepackage{simpleConference}
\usepackage{times,natbib}
\usepackage{graphicx}
\usepackage{amssymb}
\usepackage{url,hyperref}
\usepackage{cite}
\usepackage{amsmath}
\usepackage{tensor}
\usepackage{booktabs}
\usepackage{cuted}

\usepackage[table]{xcolor}
\definecolor{LightGray}{gray}{0.85}
\definecolor{Cyan}{rgb}{0.88,1,1}
\definecolor{Pink}{rgb}{1.0,.88,.88}
\definecolor{Orange}{rgb}{0.95,.85,.71}

\newcommand{\cyan}{\cellcolor{Cyan}}
\newcommand{\pink}{\cellcolor{Pink}}
\newcommand{\orange}{\cellcolor{Orange}}

\begin{document}

\title{A generalized wave-vortex decomposition for rotating Boussinesq flows with arbitrary stratification}
\author{Jeffrey J. Early, Northwest Research Associates, USA \\ M. Pascale Lelong, Northwest Research Associates, USA  \\  Miles A. Sundermeyer, University of Massachusetts Dartmouth, USA}
\maketitle
\thispagestyle{empty}

\abstract{The linear wave and geostrophic (vortex) solutions are shown to be a complete basis for physical variables $(u,v,w,\rho)$ in a rotating non-hydrostatic Boussinesq model with arbitrary stratification. As a consequence, the fluid can be unambiguously separated into linear wave and geostrophic components at each instant in time, without the need for temporal filtering. The fluid can then be diagnosed for temporal changes in wave and geostrophic coefficients at each unique wavenumber and mode, including those that inevitably occur due to nonlinear interactions.

We demonstrate that this methodology can be used to determine which physical interactions cause the transfer of energy between modes by projecting the nonlinear equations of motion onto the wave-vortex basis. In the particular example given, we show that an eddy in geostrophic balance superimposed with inertial oscillations at the surface transfers energy from the inertial oscillations to internal gravity wave modes. This approach can be applied more generally to determine which mechanisms are involved in energy transfers between wave and vortices, including their respective scales.

Finally, we show that the nonlinear equations of motion expressed in a wave-vortex basis are computationally efficient for certain problems. In cases where stratification profiles vary strongly with depth, this approach may be an attractive alternative to traditional spectral models for rotating Boussinesq flow.}
\\ \\
\noindent This work has not yet been peer-reviewed and is provided by the contributing author(s) as a means to ensure timely dissemination of scholarly and technical work on a noncommercial basis. Copyright and all rights therein are maintained by the author(s) or by other copyright owners. It is understood that all persons copying this information will adhere to the terms and constraints invoked by each author's copyright. This work may not be reposted without explicit permission of the copyright owner.


\section{Introduction}

The linearized rotating Boussinesq equations admit two solution types---inertia-gravity waves and geostrophic (vortex) motions. The wave and geostrophic solutions form the foundation for how we understand and interpret ocean and atmosphere observations in a wide variety of contexts. These two types of linear motion are predictive in the sense that knowledge of the solution at one time enables knowledge of the solution for all time.  Such solutions thus guide our intuition for how the ocean evolves, while deviations from those predictions also serve as a direct measurement of nonlinearity. For these reasons, a great deal of effort goes into separating fluid motions into these two types of solutions.

Wave and vortex solutions can be separated in the frequency-wavenumber domain by utilizing the dispersion relation of the linear wave solutions, a method well suited to model output \citep[e.g.][]{savage2017-jgr,torres2018-jgr}. With more sparse \emph{in situ} observations, other methods have been developed to make this same separation; however these typically require additional statistical assumptions to overcome the limitations of sparse sampling \citep{lien1992-jpo,buhler2014-jfm,lien2019-jpo}. In idealized Boussinesq models with triply periodic domains and constant stratification, such a decomposition can be made unambiguously at each instant in time \citep{bartello1995-jas,smith2002-jfm,waite2006b-jfm}. For each resolved wavenumber the decomposition splits the flow into two inertia-gravity waves ($A_\pm$), with frequencies that lie between the Coriolis, $f_0$, and the buoyancy frequency, $N$; and a zero-frequency, geostrophic solution ($A_0$), which accounts for all linear potential vorticity (PV). Thus, the wave-vortex decomposition is a linear transformation that projects the variables $(u,v,\rho)$ onto an equivalent representation $(A_+, A_-, A_0)$ of two wave and one vortex mode without loss of information. Note, because the transformation uses vertical eigenmodes that guarantee the continuity equation is satisfied, the vertical velocity, $w$, is redundant and not needed in the transformation. Furthermore, the inverse transformation can recover both $w$ and pressure from the wave-vortex components $(A_+, A_-, A_0)$.

Aside from being an alternative and compact representation of the dynamical variables, the wave-vortex decomposition has a number of applications. Unlike other spectral representations of fluid flow, the wave-vortex projection is useful because it projects directly onto \emph{solutions} of the equations of motion, including, as shown in this manuscript, in the case of arbitrary stratification. Coefficients of the projection are thus physically meaningful, directly encoding the amplitude and phase of the unique wave and geostrophic solutions. For example, applying the wave-vortex projection to output from a perfectly linear wave model will show no changes in amplitude and phase over time, while in contrast, a non-linear wavelike process will have amplitude and phase that become decorrelated with time. Alternatively, for flows that bear no resemblance to the linear solutions, the projection may not be meaningful---thus the interpretation of the components as representations of wave and geostrophic components is ultimately problem-specific.

One of the simplest diagnostics utilizing the wave-vortex decomposition is assessing how total energy shifts between inertial, internal gravity waves and geostrophic solutions. The physical mechanisms that transfer energy between wave-vortex modes can further be diagnosed by projecting the nonlinear equations of motion into wave-vortex space. This approach was used by \citet{lelong91-jfm,bartello1995-jas} to diagnose transfer in the turbulence cascade, and also, for example, by \citet{arbic2012-jpo} to diagnose energy transfers across modes and wavenumbers in quasigeotrophic turbulence. 
With the nonlinear equations of motion projected onto the wave-vortex modes, it is also possible to create a series of reduced-interactions models, as has been done for triply-periodic domains \citep{remmel2010-thesis,remmel2010-cms,hernandez-duenas2014-jfm}. These models are reduced versions of the equations of motions that restrict interactions between certain modes. For example, restricting interactions between only PV modes results in the quasigeostrophic equations, while restricting interactions between only wave-modes results in an extension of the weak wave turbulence model \citep{remmel2010-thesis}.

\begin{figure*}[h]
  \centerline{\includegraphics{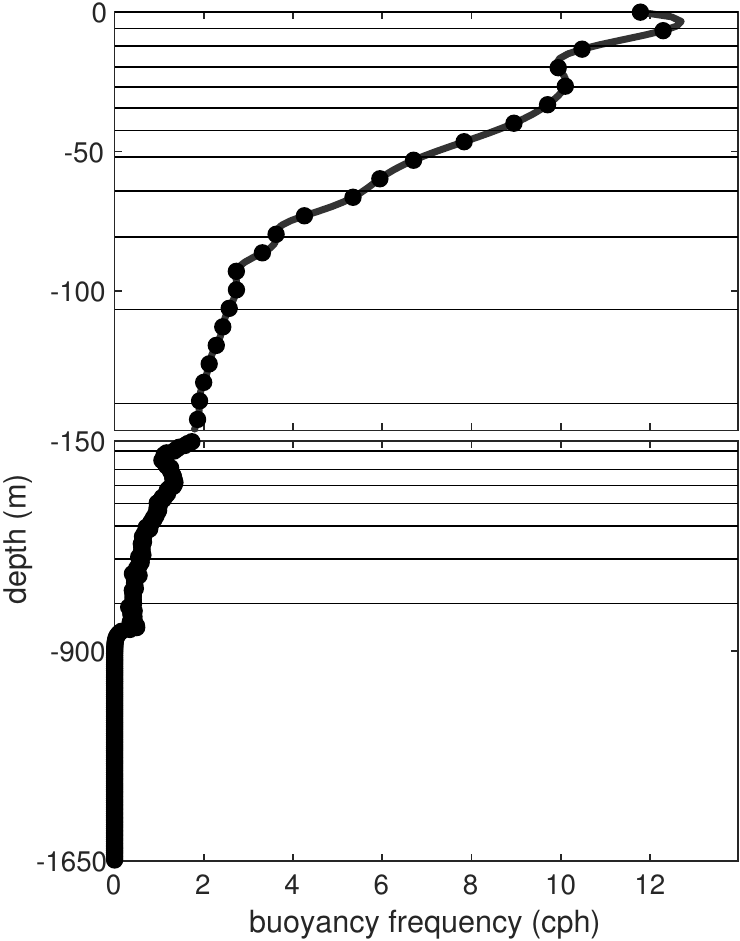}}
  \caption{Buoyancy frequency as a function of depth for a location in the Eastern Mediterranean Sea. Black dots indicate regularly-spaced grid points, while horizontal lines are the roots (Gauss quadrature points) of the 19th internal mode. Note that the vertical scale changes at 150 m depth.}
\label{fig:stratification-profile}
\end{figure*}

While the aforementioned studies have addressed and utilized the wave-vortex decomposition for the case of constant stratification, typical stratification profiles in the ocean often resemble an exponential-like function, as shown in figure \ref{fig:stratification-profile}. A computational challenge that arises when solving the equations of motion on a regular grid with such stratification is that, while the grid resolution (black dots in the figure) is more than adequate at depth, rapid variations near the surface are not resolved, even with large numbers of grid points (257 in this case).
To address this limitation, in this manuscript we extend the wave-vortex decomposition for Boussinesq flows to arbitrary non-constant stratification. This involves solving an eigenvalue problem (EVP) to obtain the vertical dependence \citep{early2020-james}, rather than Fourier mode expansions in the three spatial directions as in the case of constant stratification. We begin in Sections 2 and 3 with the linearized equations of motion and their solutions. Section 4 then details the projection onto the vertical modes, while Section 5 shows the decomposition itself. 

In Section \ref{sec:applications} we provide an example application in which we diagnose the results of the Cyprus eddy studied by \citet{lelong2020-jpo}, in which an eddy in geostrophic balance superimposed with inertial oscillations at the surface transfers energy from the inertial oscillations to generate internal gravity waves. The present decomposition is performed at each instant in time using the same model output, and is shown to agree with the temporal filtering based method used in the original study. Results of the present analysis show that advection of geostrophic vorticity by the inertial oscillations accounts for \emph{all} the energy transfer from the inertial oscillations to internal gravity waves, confirming the hypothesis in the original study.

Section \ref{sec:modeling} discusses the implications of these results and addresses general challenges encountered in numerical modeling, including the important implications that, in cases of variable stratification, regularly spaced grids may only resolve a small fraction of the physically relevant vertical modes, and as such, it may be more computationally efficient to integrate the nonlinear equations of motion in wave-vortex space than in physical space.  Finally, Section \ref{sec:conclusion} offers some concluding remarks. Appendix A shows how the results are simplified for constant stratification, and Appendix B details the numerical implementation. The projection of the nonlinear equations of motion onto the wave-vortex modes is documented in Appendix \ref{sec:nonlinear-eom}.

%
\section{Background}
\label{sec:background}
%

The linearized, unforced, inviscid equations of motion for fluid velocity $u(x,y,z,t)$, $v(x,y,z,t)$, $w(x,y,z,t)$, on an $f$-plane are
\begin{subequations}{}
\begin{align}
\label{x-momentum}
\partial_t u - f_0 v =& - \frac{1}{\rho_0} \partial_x p \\ \label{y-momentum}
\partial_t v + f_0 u =& - \frac{1}{\rho_0} \partial_y p \\ \label{z-momentum}
\partial_t w =& - \frac{1}{\rho_0} \partial_z p - g \frac{\rho}{\rho_0} \\ \label{continuity}
\partial_x u + \partial_y v + \partial_z w =& 0 \\ \label{thermodynamic}
\partial_t \rho + w \partial_z \bar{\rho} =& 0.
\end{align}
\end{subequations}
Here $p(x,y,z,t)$ and $\rho(x,y,z,t)$ are perturbation pressure and density, respectively, defined such that total pressure $p_{\textrm{tot}}(x,y,z,t) = p_0(z) + p(x,y,z,t) $ and total density $\rho_{\textrm{tot}}(x,y,z,t) = \rho_0 + \bar{\rho}(z) + \rho(x,y,z,t) $ where $\partial_z p_0(z) = -g\left( \rho_0 + \bar{\rho}(z) \right)$. All variables in \eqref{x-momentum}--\eqref{thermodynamic} are functions of $x$, $y$, $z$ and $t$, except $\bar{\rho}$, which is only a function of $z$. We use the usual definition of buoyancy frequency, $N^2(z) \equiv -\frac{g}{\rho_0} \partial_z \bar{\rho}$. Throughout this manuscript we use the linear approximation to isopycnal displacement $\eta \equiv -\rho/\bar{\rho}_z$ rather than density anomaly. With this notation \eqref{thermodynamic} becomes $w=\partial_t \eta$ and \eqref{z-momentum} can be similarly rewritten.

We assume boundaries are periodic in the horizontal, $(x,y)$, and bounded in the vertical, $z$. The lower boundary is assumed flat at $z=-D$ with free-slip and $w(-D)=0$, and no density anomaly, $\rho(-D)=0$. Similarly, the upper boundary is taken to be a free-slip rigid-lid with $w(0)=0$ and also no density anomaly, $\rho(0)=0$. The lack of density anomalies at the boundaries is restrictive and will be addressed in future work.

The depth integrated energy densities of the flow are
\begin{multline}
    \textrm{HKE}\equiv \frac{1}{2}\int_{-D}^0 (u^2+v^2)\, dz,\; \\ \textrm{VKE}\equiv \frac{1}{2}\int_{-D}^0  w^2\, dz, \;  \textrm{PE} \equiv \frac{1}{2} \int_{-D}^0 N^2 \eta^2\, dz
    \end{multline}

\noindent
where $\textrm{HKE}$, $\textrm{VKE}$, and $\textrm{PE}$ are the horizontal kinetic energy, vertical kinetic energy and potential energy per unit mass, respectively. The other conserved quantity of interest is the quasi-geostrophic potential vorticity (PV),
\begin{equation}
\label{linear-pv}
    \textrm{PV} \equiv \partial_x v - \partial_y u - f_0 \partial_z \eta
\end{equation}
\noindent
which can be directly derived from the linear equations \eqref{x-momentum}-\eqref{thermodynamic}, or found as the linear approximation to the available potential vorticity (APV) as defined by \citet{wagner2015-jfm}.

It is noteworthy that linearized Ertel PV does not correspond to a useful quantity in this model -- it is neither conserved nor time-independent for the internal gravity wave solutions (see \eqref{ertel-pv-igw}). Linearized Ertel PV is
\begin{align}
    \textrm{Ertel PV} \equiv& \frac{\left( \vec{\zeta} + \hat{k}f_0\right) \cdot \nabla \rho_\textrm{tot}}{\rho_\textrm{tot}} \\ \label{linear-ertel-pv}
    \approx & \frac{1}{\rho_0} \left(\zeta^z \bar{\rho}_z + f_0 \rho_z  + f_0 \bar{\rho}_z  \right)
\end{align}{}
\noindent
where $\vec{\zeta}$ is vorticity and $\zeta^z=\partial_x v - \partial_y u$ is its vertical component. Notable is that linear Ertel PV per \eqref{linear-ertel-pv} does not equal the conserved PV quantity \eqref{linear-pv}---the primary difficulty being that $\partial_z \eta$ is not proportional to $\partial_z \rho / \bar{\rho}_z$ for non-constant stratification. Applying the total derivative to \eqref{linear-ertel-pv} results in
\begin{align}
\label{advection-of-linear-pv}
    \frac{d}{dt} \left( \textrm{Ertel PV} \right) \approx &  \frac{1}{\rho_0} \left( \bar{\rho}_z \partial_t \zeta^z  + f_0 \partial_t \rho_z  + f_0 w \bar{\rho}_{zz}  \right)
\end{align}{}
which is not a conservation equation, but a balance between three terms: local changes in the vertical component of vorticity, local changes in the vertical gradient of the density anomaly, and the vertical advection of the background density gradient. The connection between \eqref{advection-of-linear-pv} and the conserved PV \eqref{linear-pv} is found using the thermodynamic equation \eqref{thermodynamic} and re-arranging, which reproduces equation \eqref{linear-pv},
\begin{align}
    \frac{d}{dt} \left( \textrm{Ertel PV} \right) \approx &  \frac{\bar{\rho}_z }{\rho_0} \frac{\partial}{\partial t}\left( \zeta^z  + f_0 \frac{\partial}{\partial z} \left( \frac{\rho}{\bar{\rho}_z}\right)    \right)
\end{align}{}
up to a scaling factor. The key difference between quasigeostrophic PV \eqref{linear-pv} and linear Ertel PV \eqref{linear-ertel-pv} is that the latter neglects vertical advection of the background density gradient. In the present context then, APV as defined in \citet{wagner2015-jfm} is the relevant conserved quantity.

%
\section{Wave-vortex solutions}
\label{sec:solutions}
%

Solutions to \eqref{x-momentum}-\eqref{thermodynamic} are assumed to take the separable form
\begin{equation}
\label{f_eqn_form}
    f(x,y,z,t) =  \sum_{jkl} \frac{1}{2} \tilde{f}_{jkl}(t) e^{i (k x + ly )} F_{jkl}(z) + c.c. 
\end{equation}
for $u$, $v$, $p$, and 
\begin{equation}
\label{g_eqn_form}
    g(x,y,z,t) =\sum_{jkl} \frac{1}{2} \tilde{g}_{jkl}(t) e^{i (k x + ly )} G_{jkl}(z) + c.c. 
\end{equation}
for $w$ and $\eta$. This presumes a Fourier basis satisfying the periodic boundary conditions in $x$, $y$, and real-valued functions $F_{jkl}(z)$, $G_{jkl}(z)$ satisfying the vertical boundary problem. The summation is over all wavenumbers $k$, $l$, but also over eigenmodes $j$ from the bases $\{ F_{jkl}(z) \}$ and $\{ G_{jkl}(z) \}$, indicated with subscripts to emphasize their dependence on wavenumbers $k$ and $l$. The coefficients $\tilde{f}_{jkl}(t)$, $\tilde{g}_{jkl}(t)$ are complex, encoding both amplitude and phase. The ``c.c.'' refers to the complex conjugate, which contains half the power of the real valued solutions, but no new information. Although the wave-vortex decomposition is performed at fixed time in the time domain $\tilde{f}_{jkl}(t)$, it is  useful to express solutions in the frequency domain, in which case we denote the variables with $\hat{\cdot}$, e.g., $\tilde{f}_{jkl}(t)= \hat{f}_{jkl}e^{i\omega t}$. Finally, we will often drop the subscripts ${jkl}$ entirely, and simply work with the coefficients at a fixed $j$, $k$, and $l$.

Using the thermodynamic equation \eqref{thermodynamic} to replace $w$ with $\eta_t$, solutions to \eqref{x-momentum}--\eqref{continuity} must satisfy
\begin{subequations}{}
\begin{align}
\label{x-momentum-spectral}
    i \omega \hat{u} - f_0 \hat{v} =& - i k \frac{\hat{p}}{\rho_0} \\ \label{y-momentum-spectral}
    i \omega \hat{v} + f_0 \hat{u} =& - i l \frac{\hat{p}}{\rho_0} \\ \label{z-momentum-spectral}
    \left(N^2 - \omega^2 \right) \hat{\eta} G =& -\frac{\hat{p}}{\rho_0} \partial_z F \\ \label{continuity-spectral}
    \left( i k \hat{u} + i l \hat{v}\right) F  =& -i \omega \hat{\eta} \partial_z G.
\end{align}
\end{subequations}
We now examine all possible solutions to these equations of motion by considering zero and nonzero frequency, as well as zero and nonzero horizontal wavenumbers. The vertical structure is treated in detail for each solution.

%
\subsection*{Geostrophic solutions, $\omega = 0, k^2 + l^2 = 0$}
%

Geostrophic solutions require a horizontal density anomaly, and because there is no mean vertical or horizontal density anomaly by definition, there are no mean geostrophic currents in this model. Any mean density anomaly should be subsumed into the definition of the mean density $\bar{\rho}$.

%
\subsection*{Geostrophic solutions, $\omega = 0, k^2 + l^2 > 0$}
%

Geostrophic solutions have no time variation, and the thermodynamic equation therefore implies that $w=0$. Assuming non-zero horizontal wavenumber $k^2  + l^2 > 0$, the equations of motion \eqref{x-momentum-spectral}-\eqref{continuity-spectral} reduce to
\begin{subequations}{}
\label{geostrophic_equations}
\begin{align}
\label{x-momentum-geostrophic}
    - f_0 \hat{v} =& - i k \frac{\hat{p}}{\rho_0} \\
    \label{y-momentum-geostrophic}
    f_0 \hat{u} =& - i l \frac{\hat{p}}{\rho_0} \\ \label{z-momentum-geostrophic}
    N^2 \hat{\eta} G_g =& - \frac{\hat{p}}{\rho_0} \partial_z F_g \\ \label{continuity-geostrophic}
    \left( i k \hat{u} + i l \hat{v}\right) F_g  =& 0
\end{align}
\end{subequations}
where $F_g(z)$, $G_g(z)$ denote the geostrophic vertical structure functions. The only equation of consequence for the vertical structure is \eqref{z-momentum-geostrophic}. With no vertical velocity, the rigid lid boundary conditions place no constraint on $F_g(z)$, $G_g(z)$. The decision to disallow density anomalies at the boundaries implies that $G_g(z)$ is an odd function, and therefore $F_g(z)$ is an even function. Although gravity $g$ does not enter into \eqref{geostrophic_equations} without a free surface, it is still convenient to set the separation constant in \eqref{z-momentum-geostrophic} such that $\hat{p}=\rho_0 g \hat{\eta}$ and $N^2 G_g = -g \partial_z F_g$, with $G_g(0)=G_g(-D)=0$ at the boundaries. This allows the amplitude of the solution to be expressed in terms of sea-surface height, analogous to typical notation for geostrophic motions.

The geostrophic solution, or vortex solution, is given by,
\begin{equation}
\label{geostrophic_solution}
\left[\begin{array}{c} u_g \\ v_g \\  \eta_g  \end{array}\right] =
\frac{\hat{A}_0}{2} \left[\begin{array}{c}
	-i \frac{g}{f_0} l F_g(z) \\
	i \frac{g}{f_0} k F_g(z) \\
	 G_g(z)
 \end{array}\right] e^{i\theta_0} + c.c.
\end{equation}
where $\hat{A}_0$ is a complex valued amplitude containing the phase information, and $\theta_0=k x + l y$.


As a consequence of only having one constraint connecting $F_g(z)$ and $G_g(z)$, there is no preferred set of vertical basis functions for the geostrophic solution. Any complete basis satisfying the boundary conditions can be used to represent the geostrophic solution. However, near-geostrophic theories with a different choice of scalings, such as quasi-geostrophy \citep[QG; e.g., see][]{pedlosky1987-book}, have nonzero vertical velocities and therefore still require that three-dimensional continuity be satisfied.
To maintain continuity we take \eqref{continuity-spectral} and set the separation constant to $h$, such that $F(z) = h \partial_z G(z)$ for all $z$. This additional requirement, combined with the hydrostatic vertical momentum condition $N^2 G = -g \partial_z F$, results in two Sturm-Liouville eigenvalue problems for hydrostatic (HS) vertical modes,
 \begin{equation}
     \label{qg-evp}
     \frac{d^2 G^\textrm{HS}_j}{dz^2}  =  - \frac{N^2}{gh_j} G^\textrm{HS}_j
 \end{equation}
 with boundary conditions $G^\textrm{HS}(0)=G^\textrm{HS}(-D)=0$ or,
  \begin{equation}
     \label{qg-g-evp}
     \frac{d}{dz} \left( \frac{1}{N^2} \frac{d F^\textrm{HS}_j}{dz} \right)  =  - \frac{1}{gh_j} F^\textrm{HS}_j
 \end{equation}
 with $\partial_z F^\textrm{HS}(0)=\partial_z F^\textrm{HS}(-D)=0$ where $j$ is the mode number and eigenvalue $h_j$ is the equivalent depth. It follows directly from Sturm-Liouville theory that the vertical modes resulting from the HS EVPs satisfy the orthogonality conditions
\begin{equation}
\label{hydrostatic_G_eqn}
\frac{1}{g} \int_{-D}^{0} N^2(z) G^\textrm{HS}_i G^\textrm{HS}_j \, dz = \delta_{ij},
\end{equation}
and
\begin{equation}
\label{hydrostatic_F_eqn}
\int_{-D}^0  F^\textrm{HS}_i F^\textrm{HS}_j \, dz = h_i \delta_{ij}
\end{equation}
where we have implicitly normalized the amplitude of the modes. The $\frac{1}{g}$ normalization in \eqref{hydrostatic_G_eqn} arises naturally when using a free-surface boundary condition, and is kept here for consistency.

The importance of the $\{ G^\textrm{HS}_j(z) \}$ and $\{ F^\textrm{HS}_j(z) \}$ bases are twofold. First, Sturm-Liouville theory guarantees that they are complete, and therefore capable of representing any function with the caveat that the $\{ G^\textrm{HS}_j(z) \}$ basis cannot capture nonzero boundary conditions. Second, the specific relationship between these modes is such that both continuity and the linearized vertical momentum equation are satisfied. In practice, this means that they often reflect the vertical structure of various linear solutions. It is in this sense that $\{ G^\textrm{HS}_j(z) \}$ and $\{ F^\textrm{HS}_j(z) \}$ are `preferred' bases for representing certain flows, including quasigeostrophy and hydrostatic linear internal waves.

The horizontal kinetic energy and potential energy of the geostrophic solution \eqref{geostrophic_solution} as a function of depth are found by averaging over time and horizontally, including the energy from the complex conjugate,
\begin{align}
    \textrm{HKE}_g=&\frac{\hat{A}_0^2}{4} \frac{g^2}{f_0^2}K^2 \int_{-D}^0 F_g^2(z) \, dz\\
    \textrm{PE}_g=&\frac{\hat{A}_0^2}{4} \int_{-D}^0 N^2(z) G_g^2(z)\, dz,
\end{align}{}
\noindent
where $K^2=k^2+l^2$. Vertical kinetic energy is identically zero. If we use the hydrostatic normal modes $F^\textrm{HS}_j$, $G^\textrm{HS}_j$ then depth-integrated horizontal kinetic energy reduces to $\textrm{HKE}_g=\frac{\hat{A}_0^2}{4} \frac{g^2 h_j}{f_0^2}K^2$ and depth-integrated potential energy reduces to $\textrm{PE}_g = \frac{\hat{A}_0^2}{4} g$.

The linearized potential vorticity is,
\begin{small}
\begin{multline}
    \textrm{PV}_g= -\frac{\hat{A}_0 g}{2 f_0 } \left( K^2F_g(z) - \frac{d}{dz} \left( \frac{f_0^2}{N^2} \frac{d F_g(z)}{dz} \right)  \right)  e^{i\theta_0}  \\+ c.c.
\end{multline}{}
\end{small}
as is traditionally written, or simply
\begin{equation}
    \textrm{PV}_g = -\frac{\hat{A}_0}{2 f_0 h} \left( ghK^2 + f_0^2 \right) F_g(z) e^{i\theta_0} + c.c.
\end{equation}{}

\noindent
after using \eqref{hydrostatic_F_eqn} to rewrite $F_g$. These expressions are exactly the potential vorticity identified in the quasi-geostrophic potential vorticity equation. In contrast, the Ertel PV is,
\begin{small}
\begin{multline}
    \textrm{Ertel PV}_g= \\  \frac{\bar{\rho}_z}{\rho_0} \left[ \textrm{PV}_g - \frac{\hat{A}_0 f_0}{2} \left( \partial_z \ln \bar{\rho}_z \right) G(z) e^{i \theta_0} + f_0 \right] + c.c.
\end{multline}{}
\end{small}
which does not correctly account for changes in the density gradient \citep[see also][]{wagner2015-jfm}.


Under rigid lid conditions, there also exists a barotropic mode ($j=0$) where $F^\textrm{HS}_0(z)=\textrm{const}$ with no associated buoyancy anomaly, $G^\textrm{HS}_0(z)=0$. This case will be handled separately in the decomposition.

%
\subsection*{Inertial oscillation solution, $\omega \neq 0, k^2 + l^2 = 0$}
\label{sec:io_solutions}
%

This solution has no vertical velocity, density anomaly, or pressure gradients. It is simply a horizontally uniform oscillating horizontal velocity field, with no constraints on vertical structure other than the boundary conditions. In the triply periodic model used in \citet{smith2002-jfm} this solution is referred to as the vertically sheared horizontal mode (VSHM), while in the bounded domain it is identified as the inertial oscillation solution,
\begin{equation}
\label{inertial_oscillation_arbitrary_structure}
\left[\begin{array}{c} u_I \\ v_I \\ \eta_I  \end{array}\right] =
 \left[\begin{array}{c}
	U_I \cos(f_0 t + \phi_0) F_I(z) \\
	-U_I \sin(f_0 t + \phi_0) F_I(z) \\
	0
 \end{array}\right].
\end{equation}
Here, since there is no conjugate to $k^2+l^2=0$, the amplitude is purely real. $F_I(z)$ is an arbitrary function, and can be expanded in any complete basis.  This is noteworthy because it essentially leaves the boundary conditions for $F_I(z)$ unspecified, and unlike other solutions considered here, $\partial_z F_I(0)$ and $\partial_z F_I(-D)$ are not necessarily zero. Therefore one must be careful not to expand $F_I(z)$ in a basis with unnecessarily restrictive boundary conditions. That said, there is not necessarily any physical insight to be gained from this additional freedom at the boundaries, and it would certainly be reasonable to restrict the model to solutions where $\partial_z F_I(0)=\partial_z F_I(-D)=0$. 

%
\subsection*{Wave solutions, $\omega \neq 0, k^2 + l^2 > 0$}
\label{sec:wave_solutions}
%

Similar to the geostrophic solution where we assumed that $\hat{p}=\rho_0 g \hat{\eta}$, the vertical momentum equation requires that $(N^2-\omega^2)G = -g \partial_z F$. Combined with continuity $F = h \partial_z G$, the vertical dependence vanishes from the problem and we are left with
\begin{equation}
    \left[\begin{array}{ccc}
    i \omega & -f_0 &  i g k \\
    f_0 & i \omega &  i g l \\
    k h & l h &  \omega 
    \end{array}\right]
    \left[\begin{array}{c}
    \hat{u} \\
    \hat{v} \\
    \hat{\eta}
    \end{array}\right]
    =
    \left[\begin{array}{c}
    0 \\
    0 \\
    0
    \end{array}\right].
\end{equation}{}

\noindent
This system of equations admits the internal wave solutions when
\begin{equation}
\label{dispersion_relation}
\omega = \sqrt{g h K^2 + f_0^2}.
\end{equation}
 The $\pm$ wave solutions are given by,
\begin{equation}
\label{wave_solutions}
\left[\begin{array}{c} u_\pm \\ v_\pm \\ \eta_\pm  \end{array}\right]  = \frac{\hat{A}_\pm}{2}
\left[\begin{array}{c} \frac{k\omega \mp il f_0}{\omega K} F(z)  \\
\frac{l \omega \pm i k f_0}{\omega K } F(z) \\
\mp \frac{K h}{\omega} G(z) \end{array}\right] e^{i \theta_\pm} + c.c.
\end{equation}
where the horizontal phase is given by $\theta_\pm=k x + l y \pm \omega t + \phi$ and the amplitude is chosen so that depth-integrated total energy is $\hat{A}^2 h/2$, as will be shown below.

Combining the vertical constraints from non-hydrostatic vertical momentum $(N^2-\omega^2)G = -g \partial_z F$ and continuity $F = h \partial_z G$ with the dispersion relation \eqref{dispersion_relation} results in the $K$-constant, non-hydrostatic Sturm-Liouville problem \citep{early2020-james},
\begin{equation}
\label{vertical-eigenvalue-G-with-K}
\frac{d^2 G_j}{dz^2} - K^2 G_j = -\frac{N^2 - f_0^2}{g h_j }G_j.
\end{equation}
The eigendepth $h_j$ and eigenfrequency $\omega_j$ are interchangeable using the dispersion relation \eqref{dispersion_relation} with fixed $K$. Note that the EVP could have been written in terms of a fixed frequency $\omega$ (with no subscript $j$), with eigendepth $h_j$ and eigenwavenumber $K_j$ (with subscript $j$), but the constant frequency formulation is not relevant for the decomposition problem at fixed time.

The depth integrated energies for the $j$-th internal wave mode at total wavenumber $K$ are,
\begin{align}
    \label{hke}
HKE_\pm =& \frac{\hat{A}_\pm^2}{4}  \left( 1+ \frac{f_0^2}{\omega_j^2} \right)  \int_{-D}^0 F_j^2(z) \, dz \\ \label{vke}
VKE_\pm =& \frac{\hat{A}_\pm^2}{4}  K^2 h_j^2  \int_{-D}^0 G_j^2(z) \, dz \\ \label{pe}
PE_\pm =& \frac{\hat{A}_\pm^2}{4}   \frac{K^2 h_j^2}{\omega_j^2}  \int_{-D}^0 N^2(z) G_j^2(z) \, dz.
\end{align}{}

\noindent
which sum to a depth-integrated total energy of $\frac{\hat{A}_\pm^2 h_j}{2}$. The internal wave solutions have zero potential vorticity per \eqref{linear-pv}, PV$_\pm=0$; but they do have Ertel PV per \eqref{linear-ertel-pv},
\begin{small}
\begin{multline}
\label{ertel-pv-igw}
    \textrm{Ertel PV}_\pm = \\  \frac{\bar{\rho}_z}{\rho_0} \left[ \pm \frac{\hat{A}_\pm}{2 } \frac{K h_j f_0}{\omega_j} ( \partial_z \ln \bar{\rho}_z ) G(z) e^{i \theta_\pm} + f_0 \right]  + c.c.
\end{multline}{}
\end{small}
again suggesting that Ertel PV may not be the appropriate quantity for this model.

%
\section{Orthogonality and projection}
\label{sec:vertical_modes}
%

The primary challenge that separates this wave-vortex decomposition from previous ones is dealing with the vertical modes resulting from the $K$-constant EVP in \eqref{vertical-eigenvalue-G-with-K}. In a vertically periodic domain with constant stratification in $z$, Fourier series are an appropriate basis. For a vertically bounded domain with arbitrary stratification in $z$ and no buoyancy anomaly at the boundaries, the appropriate basis are the eigenmodes $G_j$ of \eqref{vertical-eigenvalue-G-with-K} with $G(0)=G(-D)=0$.

%
\subsection*{Orthogonality}
\label{sec:orthogonality}
%

The non-hydrostatic Sturm-Liouville problem given by \eqref{vertical-eigenvalue-G-with-K} implies that for a given wavenumber $K$, two modes $G_i(z)$, $G_j(z)$ satisfy the orthogonality condition,
\begin{equation}
\label{k_const_ortho}
\frac{1}{g} \int_{-D}^{0} (N^2(z)-f_0^2) G_i G_j \, dz = \delta_{ij}
\end{equation}
where we have normalized the modes. Unlike the hydrostatic case, there does not appear to be an equivalent Sturm-Liouville problem for the non-hydrostatic $F_j$ modes (with constant $K$) and therefore no associated orthogonality condition. The expression
\begin{equation}
    \partial_z \left( \frac{\partial_z F_j}{N^2 - g h_j K^2 - f_0^2} \right) = -\frac{1}{g h_j} F_j,
\end{equation}
as far as we know, cannot be coerced to Sturm-Liouville form. The closest relationship we are able to find is
\begin{equation}
\label{k_const_F_ortho}
\int_{-D}^0  \left( F_i F_j + h_i h_j K^2 G_i G_j \right) \, dz = h_i \delta_{ij}.
\end{equation}
The difference between \eqref{hydrostatic_F_eqn} and \eqref{k_const_F_ortho} is significant -- the former can be used on any function satisfying the boundary conditions, while the latter requires a specific relationship between the dynamical variables to project on the $F_j$ modes.

%
\subsection*{Projection}
\label{sec:projection}
%

If a dynamical variable that expands in $G$, such as density anomaly, $\rho(z)$, satisfies the appropriate boundary conditions, it can be written as in \eqref{g_eqn_form}, e.g.,
\begin{equation}
    \rho(x,y,z,t) =  \sum_{jkl}\frac{1}{2} \tilde{\rho}_{jkl}(t) G_{jlk}(z) e^{i(kx+ly)} + c.c.
\end{equation}{}


\noindent 
where the coefficients are recovered with
\begin{multline}
\label{projection}
\tilde{\rho}_{jkl}(t) = \frac{1}{g} \int_{-D}^0 (N^2(z) - f_0^2)\\ \cdot \left[ \frac{1}{N_x N_y} \sum_{xy} \rho(x,y,z,t) e^{-i(kx+ly)} \right] G_{jkl}(z) \, dz.
\end{multline}{}

\noindent
The projection operation \eqref{projection} first requires taking a Fourier transform of the variable, then invoking the orthogonality condition \eqref{k_const_ortho} with $j$-th vertical mode $ G_{jkl}(z)$ for wavenumber $K=\sqrt{k^2+l^2}$. However, in order to use orthogonality condition \eqref{k_const_F_ortho} as a projection operator, dynamical variables expanded in $F$ must be added to a related dynamical variable that scales like $h G$. For example, the divergence, $\delta=\partial_x u + \partial_y v$, and vertical vorticity, $\zeta=\partial_x v - \partial_y u$, can be recovered from the wave solution \eqref{wave_solutions} with,
\begin{align}
    \delta_j(t) =& \int_{-D}^0 \left( \delta(t) F_j(z) - i K^2 w(t) h_j G_j(z) \right) \, dz \\
    \zeta_j(t) =& \int_{-D}^0 \left( \zeta(t) F_j(z) - i f_0 K^2 \eta(t) h_j G_j(z) \right) \, dz
\end{align}{}

\noindent
where $\delta(z,t) = \sum \delta_j(t) F_j(z)$ and $ \zeta(z,t) = \sum \zeta_j(t) F_j(z)$. However, this only works for wave solutions since the geostrophic solution does not have the same relationships between $(u,v)$ and $\eta$. It thus appears that \eqref{k_const_F_ortho} is not particularly useful in recovering solutions.

To project variables $u$ and $v$ (and also $p$) that are expanded in $F$, we instead use the relationship derived from continuity, $F_j(z)=h_j \partial_z G_j(z)$, and consider the depth-integrated quantities. That is, if
\begin{equation}
    \label{u-xyzt}
    u(x,y,z,t) = \sum_{j,k,l} \frac{\tilde{u}_{jkl}(t)}{2} e^{i (k x + ly)} F_{jkl}(z) + c.c., 
\end{equation}
then we compute $U=\int_{-D}^z u \, dz^\prime$ so that,
\begin{small}
\begin{equation}
    \label{U-xyzt}
    U(x,y,z,t) = \sum_{j,k,l} \frac{\tilde{u}_{jkl}(t)}{2} e^{i (k x + ly)} h_{ijk} G_{jkl}(z)  + c.c.,
\end{equation}{}
\end{small}
\noindent
which can then be projected using \eqref{projection} to recover $\tilde{u}_{jkl}(t)$. Notable here is that the depth-integrated quantities represented by \eqref{U-xyzt} are themselves depth dependent.

As discussed in the next section, the only part of the solution that must be handled in a special manner is the barotropic $j=0$ mode $F_0(z)$, which as previously discussed has no projection on the $G$ modes in the rigid lid case. In practice, the integration linking \eqref{u-xyzt} to \eqref{U-xyzt} can be performed by projecting $u$ onto either the $\{ F^\textrm{HS}_j(z) \}$ basis or a cosine basis (either of which satisfy the correct boundary conditions and have a constant/barotropic mode), integrating spectrally, and then transforming back to the spatial domain.

%
\section{Wave-vortex decomposition}
\label{sec:wave-vortex-decomposition}
%

Per the previous discussion, the wave vortex decomposition requires integrating $(u,v)$ to get $(U,V)$, taking the Fourier transform in the horizontal of $(U,V,\eta)$, and then projecting the vertical structure at each horizontal wavenumber $k$ and $l$ onto the vertical eigenmodes found via the $K$-constant EVP, \eqref{vertical-eigenvalue-G-with-K}.
 \citet{early2020-james} developed a methodology for the computation and projection onto these modes. Written as a sum of individual linear solutions, and explicitly including the dependence on $j,k,l$, the three required variables are expressed as
\begin{align}
        U(x,y,z,t) =& \sum_{j,k,l} \frac{\tilde{U}_{jkl}(t)}{2} e^{i (k x + ly)}  G_{jkl}(z) + c.c. \\
    V(x,y,z,t) =& \sum_{j,k,l} \frac{\tilde{V}_{jkl}(t)}{2} e^{i (k x + ly)} G_{jkl}(z) + c.c. \\
    \eta(x,y,z,t) =& \sum_{j,k,l} \frac{\tilde{\eta}_{jkl}(t)}{2} e^{i (k x + ly)} G_{jkl}(z)  + c.c.
\end{align}{}

\noindent
where $\tilde{U}_{ijk}(t)= \tilde{u}_{ijk}(t) h_{ijk}$ and $\tilde{V}_{ijk}(t)= \tilde{v}_{ijk}(t) h_{ijk}$. The horizontal Fourier transform followed by the vertical projection then recovers $\tilde{U}_{ijk}(t)$, $\tilde{V}_{ijk}(t)$, and $\tilde{\eta}_{ijk}(t)$.

%
\subsection*{Nonzero wavenumber solutions, $k^2 + l^2 > 0$, $j=0$}
\label{sec:nonzero_k_zero_j_inversion}
%

When vertically integrating the horizontal velocities $u$, $v$ to project onto the vertical modes, the amplitude of the $j=0$ mode must be handled separately. The $j=0$ mode for the rigid lid boundary condition has no density anomaly, $\tilde{\eta}(t)=0$, and no divergence, $\tilde{\delta}(t) = i k \tilde{u}(t) + i l \tilde{v}(t)=0$. This leaves only the amplitude and phase of the vorticity $\tilde{\zeta}(t) = i k \tilde{v}(t) - i l \tilde{u}(t)$. The only valid solution is therefore the vortex solution,
\begin{equation}
\label{vortex_inversion}
    \hat{A}_0 = - i \frac{f_0}{g K^2} \left( k \tilde{v}(t) - l \tilde{u}(t) \right)
\end{equation}{}
valid for all $k^2 + l^2 > 0$.

%
\subsection*{Nonzero wavenumber solutions, $k^2 + l^2 > 0$, $j>0$}
\label{sec:nonzero_k_inversion}
%

For each wavenumber $(k,l)$ and mode $j$ there are six unknowns: the amplitudes and phases of the three different solutions. We denote the complex amplitudes as $\hat{A}_+$, $\hat{A}_-$, and $\hat{A}_0$, for the positive and negative wave, and geostrophic solutions, respectively. In matrix form the three linearly independent solutions from \eqref{geostrophic_solution} and \eqref{wave_solutions} at wavenumbers $k$, $l$, and mode $j$ are given by
\begin{small}
\begin{equation}
\label{wavevortex-forward-transformation}
\left[\begin{array}{c}\tilde{U}(t) \\\tilde{V}(t) \\ \tilde{\eta}(t)\end{array}\right] = 
\left[\begin{array}{ccc} \frac{k\omega - il f_0}{\omega K}h & \frac{k\omega + il f_0}{\omega K}h & -i \frac{gh}{f_0} l \\
\frac{l \omega + i k f_0}{\omega K }h & \frac{l\omega - i k f_0}{\omega K }h & i\frac{gh}{f_0} k  \\
- \frac{K h}{\omega} &  \frac{K h}{\omega} & 1\end{array}\right]
\left[\begin{array}{c}e^{i \omega t} \hat{A}_+  \\ e^{-i \omega t} \hat{A}_-  \\\hat{A}_0 \end{array}\right]
\end{equation}
\end{small}
which  can be inverted to solve for $\hat{A}_+$, $\hat{A}_-$, and $\hat{A}_0$,
\begin{small}
\begin{subequations}
\label{wavevortex-inverse-transformation}
\begin{align}
    \hat{A}_\pm =& \frac{e^{\mp i\omega t}}{2} \left[ \frac{k \omega \pm i l f_0}{\omega K h} \tilde{U}(t) + \frac{l \omega \mp i k f_0}{\omega Kh}\tilde{V}(t) \mp \frac{gK}{\omega} \tilde{\eta}(t) \right] \\
    \hat{A}_0 =& i \frac{l f_0}{\omega^2} \tilde{U}(t) - i \frac{k f_0}{\omega^2}\tilde{V}(t) + \frac{ f_0^2}{ \omega^2} \tilde{\eta}(t).
\end{align}{}.
\end{subequations}
\end{small}

There is some insight to be gained by defining depth-integrated versions of horizontal divergence and potential vorticity,
\begin{multline}
\tilde{\Delta}(t) =   i \left( k \tilde{U}(t) + l \tilde{V}(t) \right), \\ \tilde{\Pi}(t) \equiv i \left( k \tilde{V}(t) - l \tilde{U}(t) \right) - f_0 \tilde{\eta}(t).
\end{multline}
Now the solution has the form,
\begin{subequations}{}
\label{wave_inversion}
\begin{align}\label{wave_coeff_soln}
\hat{A}_+ =& \frac{e^{-i \omega t}}{2 K h} \left[  - i\tilde{\Delta}(t) - \omega \left( \frac{f}{\omega^2} \tilde{\Pi}(t) + \tilde{\eta}(t) \right) \right]  \\
\hat{A}_- =& \frac{e^{i \omega t}}{2Kh} \left[  - i\tilde{\Delta}(t) + \omega \left( \frac{f}{\omega^2} \tilde{\Pi}(t) + \tilde{\eta}(t) \right) \right] \\ \label{vortex_coeff_soln}
\hat{A}_0 =& - \frac{f}{\omega^2} \tilde{\Pi}(t).
\end{align}
\end{subequations}
Importantly, \eqref{wave_coeff_soln}--\eqref{vortex_coeff_soln} show that the vortex solution is recovered directly from potential vorticity and the \emph{sum} of the two wave solutions is recovered from the divergence of the transport. Extracting the phase information and energetics of individual wave solutions still requires additional information from vorticity and isopycnal displacement.

%
\subsection*{Zero wavenumber solutions, $k^2 + l^2 = 0$}
\label{sec:zero_k_inversion}
%

The only $k^2 + l^2 = 0$ solution is still inertial oscillations, per \eqref{inertial_oscillation_arbitrary_structure}, with simple rotation and zero isopycnal displacement, i.e., 
\begin{subequations}{}
\label{inertial_inversion}
\begin{align}
    u_I(0) =& \left[\tilde{u}(t) \cos(f_0 t) - \tilde{v}(t) \sin(f_0 t)\right] F_I(z) \\
    v_I(0) =& \left[\tilde{u}(t) \sin(f_0 t) + \tilde{v}(t) \cos(f_0 t)\right] F_I(z) \\
    \eta_I(0) =& 0.
\end{align}{}
\end{subequations}

%
\subsection*{Summary of the decomposition}
\label{sec:summary}
%

A key feature of the decomposition is that the recovered coefficients \eqref{vortex_inversion}, \eqref{wave_inversion} and \eqref{inertial_inversion} are strictly independent of time when applied to time-dependent linear solutions. That is, the left-hand sides of these equations are time-independent, while the right-hand sides contain terms that are time-dependent. This is not a contradiction; it simply reflects the fact that for unforced inviscid flow, the amplitude and phase of the linear solutions will remain fixed for all time. Applying the linear decomposition to nonlinear flows, an important implication of the latter result is that the actual linearity or nonlinearity of a flow can be made precise by assessing time variation in the recovered coefficients. For example, if $\hat{A}_+$ for a given $j,k,l$ at time $t=t_0$ is exactly equal to $\hat{A}_+$ computed at time $t=t_1$, then that component of the flow was perfectly linear in the sense that the wave solution \eqref{wave_solutions} exactly described its evolution.
The key takeaway when applying the above decomposition is that \emph{any} time variation in the recovered coefficients, \eqref{vortex_inversion}, \eqref{wave_inversion} and \eqref{inertial_inversion}, \emph{by definition} implies nonlinearity.

%
\section{Nonlinear transfers between wave and vortex solutions}
\label{sec:applications}
%

Having derived a generalized wave-vortex decomposition for arbitrary stratification, we next demonstrate its utility by analyzing output from a nonlinear numerical simulation performed by \citet{lelong2020-jpo} using the Boussinesq model described by \citet{winters2012-om}. The study by \citet{lelong2020-jpo} was motivated by evidence of intense near-inertial wave activity at the base of the quasi-permanent Cyprus eddy in the Eastern Mediterranean \citep{cuypers2012-bg}, and was designed to explain the origin and dynamics of the observed waves. 

Background stratification in the region surrounding the Cyprus eddy changes rapidly near the surface, following an approximately exponential-like profile as shown in figure~\ref{fig:stratification-profile}. 
Within this stratification, the Cyprus eddy was modeled as an axisymmetric vortex in geostrophic equilibrium via the streamfunction,
\begin{equation}
    \psi = -\frac{1}{4} \frac{A}{\alpha} e^{-2 \alpha (x^2 + y^2) - 2\beta z^2},
\end{equation}
where $(u_\textrm{g},v_\textrm{g},\rho_\textrm{g}) = (-\partial_y \psi,\partial_x \psi,-\rho_0 f_0 \partial_z \psi/g)$, and the strength and extent of the eddy were set by parameters $A$, $\alpha$, and $\beta$, chosen to closely match the observations. This eddy initial condition projects exactly onto to the geostrophic solutions in Section \ref{sec:solutions}, and remains stable in the nonlinear model. To model the effects of an impulsive wind stress at the surface, superimposed on the geostrophic vortex initial condition is an inertial oscillation of the form 
\begin{equation}
    u_I = U_I e^{-\gamma z}, v_I = 0,
\end{equation}
which itself projects exactly onto the inertial solution in Section~\ref{sec:solutions}. In the absence of the eddy, the inertial oscillation also remains stable in a nonlinear $f$-plane model. However, as shown by \citet{lelong2020-jpo}, the combined presence of the anticyclonic eddy and the inertial oscillation causes the inertial oscillation to lose energy while generating slightly subinertial internal gravity waves that propagate downward into the eddy core. 

\begin{figure*}[h]
  \centerline{\includegraphics{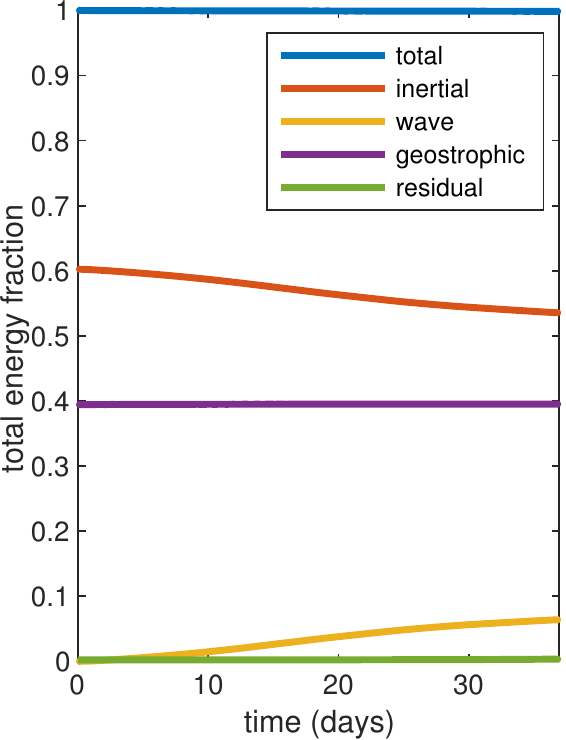}}
  \caption{Energy time series for the Cyprus Eddy simulation inferred via the wave-vortex decomposition, showing that total energy and geostrophic energy are approximately conserved, while inertial energy decreases and internal gravity wave  increases by comparable amounts.  Residual energy increases slightly (from 0.2\% to  0.3\%) during the same period.}
\label{fig:energy-fraction}
\end{figure*}

The transfer of energy from inertial oscillations to internal gravity waves was estimated using spatial-temporal averaging by \citet{lelong2020-jpo}, but can also be computed diagnostically at each instant in time using the wave-vortex decomposition described in the previous sections.
Figure \ref{fig:energy-fraction} shows energy time series for the inertial, geostrophic and wave mode solutions computed via the wave-vortex decomposition, which are consistent with the transfer of energy observed in \citet{lelong2020-jpo} (their figure 12b). 

In addition to total energies, the present methodology also enables us to examine more precisely which scales are involved in the energy transfers, and further identify exactly the dynamical mechanisms that are responsible. 
\begin{figure*}[h]
  \centerline{\includegraphics{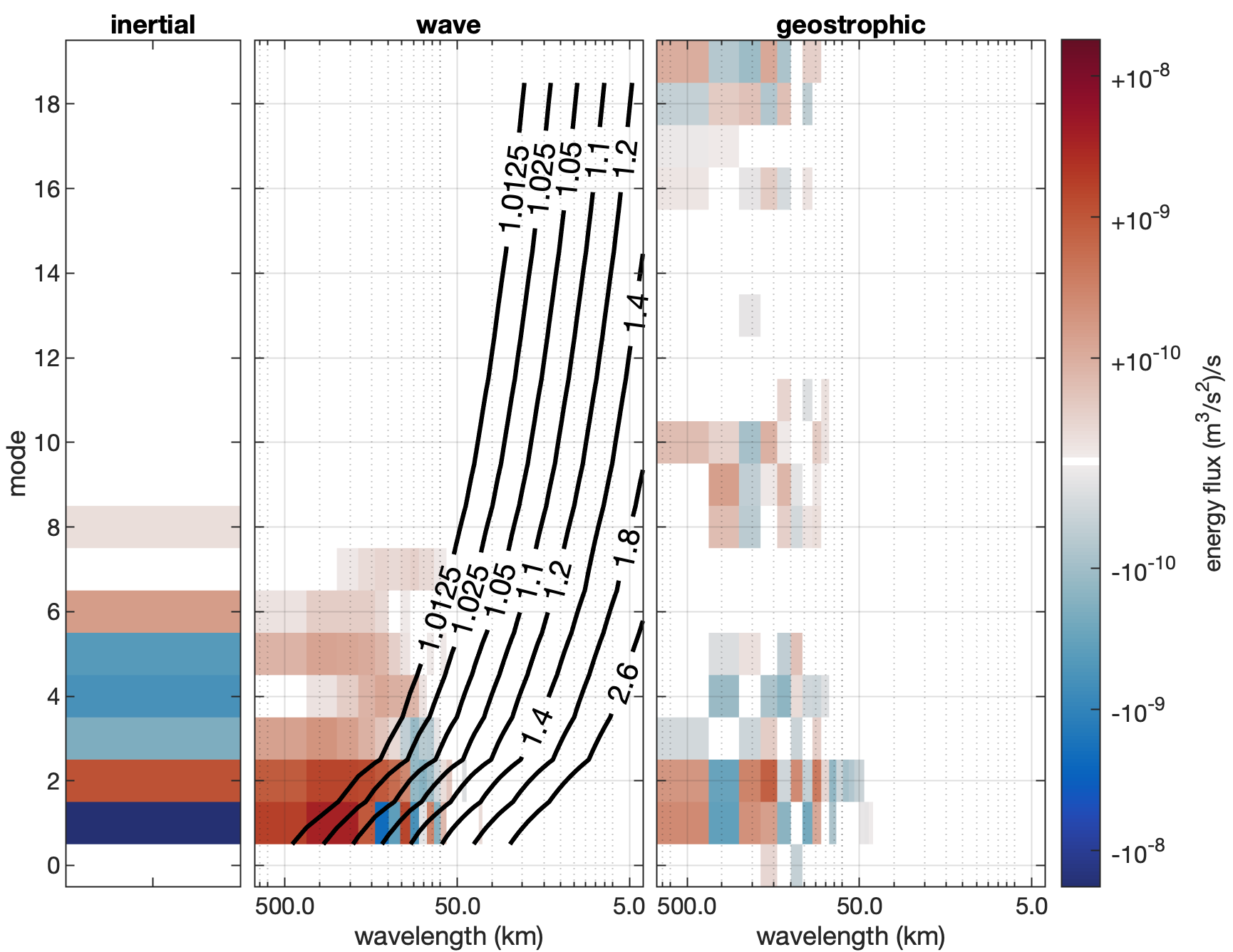}}
  \caption{Change in energy of inertial, wave and geostrophic components as a function of mode and horizontal scale on day 6 of the Cyprus eddy simulation. Contours on the wave plot indicate the frequency of oscillation, in units of $f_0$. The change in energy is dominated by a sustained loss of energy in the $j=1$ inertial mode, and a significant gain in wave energy at scales around 250 km. Changes in geostrophic energy are an order of magnitude smaller (note the log color scale), and rapidly oscillate between signs with no sustained gain or loss.  }
\label{fig:energy-flux-kj}
\end{figure*}
Figure~\ref{fig:energy-flux-kj} shows the  change in energy among the different vertical modes and horizontal scales for geostrophic, inertial, and wave components on day 6 of the simulation. The dominant energy transfer for all three components is in the lowest modes. Only the geostrophic flow shows a weak signature of energy transfer in higher modes, although at later times energy transfers occur at higher internal gravity wave modes as well (not shown). As anticipated by \citet{lelong2020-jpo}, the peak energy transfer into the waves occurs at horizontal wavelengths similar to the length scales of the geostrophic flow, with corresponding near-inertial frequencies  deviating from $f_0$ by only a few percent. Note, however, that the original study found wave frequencies to be slightly \emph{subinertial}, an effect caused by the eddy's anticyclonic geostrophic vorticity \citep{kunze1985-jpo}. This shift to subinertial frequencies is not directly captured by the linear decomposition, and instead the subinertial waves alias into other superinertial frequencies.

\citet{lelong2020-jpo} identified the vertical gradient of advection of geostrophic vorticity by inertial oscillations as the most likely dynamical mechanism for transferring energy from inertial oscillations to internal gravity waves. This is consistent with figure \ref{fig:energy-fraction}, which suggests that the waves draw their energy entirely from the inertial flow, with the geostrophic flow acting as a catalyst in facilitating energy transfer. Although the evidence presented in \citet{lelong2020-jpo} is entirely consistent with this hypothesis, the authors were not able to show conclusively whether this mechanism can account for the total energy transferred from inertial oscillations to internal gravity waves.

\begin{figure*}[h]
  \centerline{\includegraphics{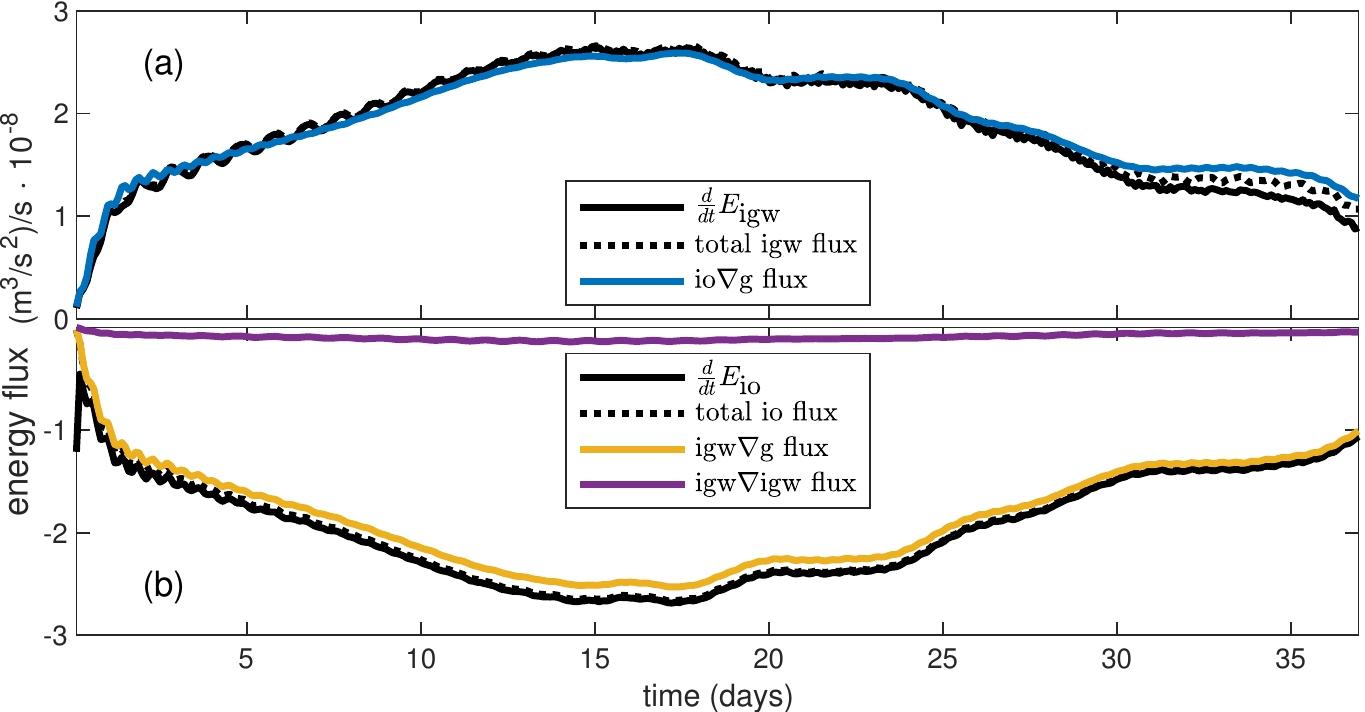}}
  \caption{Total change in energy (black) and computed total flux (dashed black) of (a) the internal gravity wave energy and (b) inertial energy. Panel (a) also shows the flux from inertial oscillations advecting geostrophic motions (blue). Panel (b) includes the flux from internal gravity waves advecting geostrophic motions (orange), and self advection of internal gravity waves (purple).}
  \label{fig:energy-flux}
\end{figure*}

To compute the energy transfers between inertial, wave, and vortex modes, we rewrite the nonlinear equations of motion by projecting them onto wave-vortex space in appendix \ref{sec:nonlinear-eom}. For the problem considered here, the internal wave frequencies do not exceed $3f_0$ and we are therefore able to make the hydrostatic approximation, simplifying the mathematics and reducing numerical complexity. Summarizing the results from the appendix, the equations of motion take the form,
\begin{equation}
\label{wave-vortex-nonlinear-eom-repeated}
    \partial_t \left[\begin{array}{c} \hat{A}_+  \\  \hat{A}_-  \\\hat{A}_0 \end{array}\right] = \left[\begin{array}{c}
 F_+ \\
 F_- \\
F_0
\end{array} \right]
\end{equation}
where the nonlinear terms are encapsulated in the three terms $F_{\pm 0}$. The transfer of energy is then proportional to $\mathcal{R}(F_{\pm 0} \bar{A}_{\pm 0})$ according to \eqref{energy-flux}. To confirm that the energy flux term is computed correctly, figure~\ref{fig:energy-flux} compares the total change in energy of the constituent parts determined via the wave-vortex decomposition at each time step, to the energy flux terms computed from $\mathcal{R}(F_{\pm 0} \bar{A}_{\pm 0})$ also computed at each time step. The two lines are nearly indiscernible, indicating that the wave-vortex projection of the nonlinear equations of motion is correctly reproducing the dynamics of the Boussinesq model. 

Appendix \ref{sec:nonlinear-eom} further shows that the nonlinear flux of energy into internal gravity waves via advection of geostrophic vorticity $\zeta_\textrm{g}$ by inertial oscillations $\left(u_\textrm{io},v_\textrm{io}\right)$ can be written as,
\begin{align}
    F_\pm^{\textrm{io}\nabla\textrm{g}} =& \pm \frac{f_0  e^{\mp i\omega t}}{2 \omega K} \left( \mathcal{F} \cdot \mathcal{DFT}\left[u_\textrm{io} \partial_x \zeta_\textrm{g} + v_\textrm{io} \partial_y \zeta_\textrm{g} \right] \right)
\end{align}
where $\mathcal{F}$ is the projection operator onto the $F_j$ modes and $\mathcal{DFT}$ is the discrete Fourier transform. Figure~\ref{fig:energy-flux}a shows that this mechanism accounts for all of the transfer of energy into internal gravity waves, directly confirming the hypothesis of \citet{lelong2020-jpo}. Additionally, figure~\ref{fig:energy-flux}b shows that the primary mechanism draining energy from inertial oscillations is the advection of the geostrophic flow by internal gravity waves, with a small contribution from self advection by internal gravity waves.

\begin{figure*}[h]
  \centerline{\includegraphics{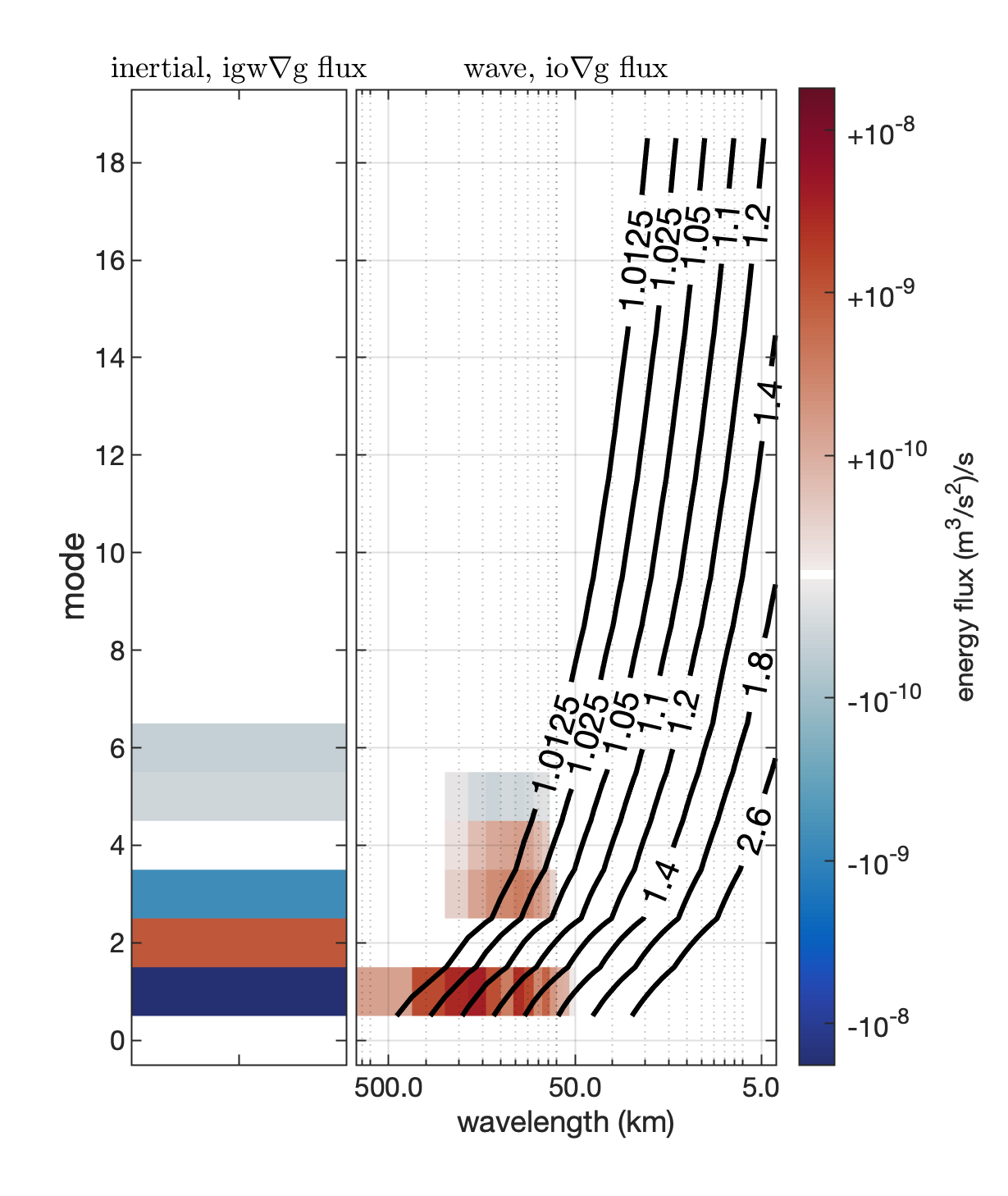}}
  \caption{Same as figure \ref{fig:energy-flux-kj}, but showing only the two dominant mechanisms from figure \ref{fig:energy-flux}. The wave panel shows flux from inertial oscillations advecting geostrophic motions and the inertial panel shows flux from internal gravity waves advecting geostrophic motions.}
\label{fig:energy-flux-kj-mechanisms}
\end{figure*}

Having identified the two dominant physical mechanisms responsible for the nonlinear transfer of energy from inertial oscillations to internal gravity waves, we can now examine exactly which modes and scales are involved in the transfer. Figure \ref{fig:energy-flux-kj-mechanisms} shows the nonlinear fluxes from only the two transfer mechanisms identified above, revealing the same dominant modes of transfer as in figure \ref{fig:energy-flux-kj}. Here it is clear that the advection of geostrophic flow by internal gravity waves primarily drains energy \emph{from} the $j=1$ inertial mode, but also shifts some energy \emph{to} the $j=2$ mode. The advection of geostrophic flow by inertial oscillations transfers energy into the mode $j=1$ internal gravity waves at scale between 50 and 500 km. The results of figure \ref{fig:energy-flux-kj-mechanisms} further imply that broader range of modes and scales seen to be exchanging energy in figure \ref{fig:energy-flux-kj} must be explained by other mechanisms that transfer energy within the internal gravity wave modes. Indeed, at later times we find that internal gravity wave energy cascades to higher modes (not shown).

%
\section{Modeling implications}
\label{sec:modeling}
%

In addition to the physical insights gained from applying the wave-vortex decomposition, there are also several implications for numerically modeling fluid flows. The first is the rather startling recognition that for the exponential stratification profile in figure \ref{fig:stratification-profile}, 257 evenly spaced vertical grid points in a pseudospectral model only resolves approximately 19 vertical modes. Conversely, this suggests that in cases of challenging stratification, modeling the equations of motion in wave-vortex space, as in appendix \ref{sec:nonlinear-eom}, may actually be more computationally efficient than traditional spectral approaches.

One of the central claims of this manuscript is that the above wave-vortex decomposition can account for all variance of $(u,v,w,\rho)$ at any instant in time. In practice, however, ocean data and numerical models have a finite number of grid points, $N$, and it is not true in general that $N$ vertical modes will be resolved with $N$ grid points. As noted in \citet{early2020-james}, only if the grid points are at (or near) the Gaussian quadrature points of the vertical modes, $F(z)$ and $G(z)$, for all resolved $K$, will all variance project onto the modes. For the case of constant stratification, the Gaussian quadrature points are evenly spaced and the vertical modes coincide with cosine and sine bases. In this special case, the $N$ vertical grid points in the Boussinesq model will coincide with $N-2$ resolved internal modes, leaving only the Nyquist frequency unresolved plus a barotropic mode. However, as figure \ref{fig:stratification-profile} demonstrates, for variable stratification, regions of rapidly changing stratification will lack resolution if an evenly spaced grid is used.

Because of the above issue, many numerical models use alternative vertical coordinates such as $\sigma$ (pressure) coordinates and density coordinates, or other more complicated hybrids, in order to better resolve the solutions. To resolve internal wave modes, \citet{early2020-james} showed that a WKB-scaled coordinate more efficiently positions grid points than a density coordinate when capturing vertical modes. Once the vertical modes are computed, the Gauss quadrature points for the first $N$ modes can computed from the roots of the $N+1$th vertical mode. When creating a numerical model, these points are the optimal choice for resolving the vertical modes.

So what happens when grid points in a numerical model are not able to fully resolve the physics? From a diagnostic point of view, any variance not captured by the $M$ resolvable modes, results in a residual. The residual of projecting a function $f$ onto the $F$ modes is defined as $f_R \equiv f-\mathcal{F}^{-1}\left[ \mathcal{F}[f] \right]$ where $\mathcal{F}$ projects using $M$ modes. For the Cyprus eddy example, the residual energy is shown in green in figure \ref{fig:energy-fraction}, and represents at most $0.3\%$ of the total energy, a $48\%$ increase from its initial value. Thus the initial conditions start with variance unresolved by the modal decomposition, and nonlinear processes further shift some of the resolved variance into unresolved variance over the course of the simulation. Because the evenly spaced $z$ grid in the model fully resolves the higher modes at lower depths, but only 19 modes near the surface, we expect an accumulation of residual energy at depth. Indeed, we find two peaks in residual energy between 450-500 m depth on day 37 of the simulation. How exactly this affects the resulting physics is not entirely clear.  All we can say is that, given higher resolution, this residual energy would be associated with higher-mode internal waves which are presently not being represented correctly. 

In the case of ocean observations with limited vertical sampling resolution, the issue is completely different than with a numerical model. In this case the physics is certainly evolving correctly, but the unresolved wave modes alias into the lower modes. Depending on the spectral slope of the unresolved modes, this aliasing may or may not have a significant impact on the coefficients of the resolved modes \citep{early2020-james}.

A second but related implication for modeling is that it may be advantageous to model the nonlinear equations of motions directly in wave-vortex space. This has the advantage of establishing which wave and vortex solutions are valid \emph{a priori}, using the methods discussed above. Additionally, damping and/or small scale variance removal is then performed directly on the wave and vortex coefficients, which correspond to wave energy and potential enstrophy damping.

In light of the effective vertical resolution issues discussed above, the computational efficiency of modeling the equations of motion in wave-vortex space is relatively favorable for cases of nonlinear stratification. The rate limiting steps are the horizontal and vertical transformations required to compute the terms in \eqref{nonlinear-rhs}.  The two basic numerical operations to be performed on a vector of length $N$ are a matrix multiplication, which requires $2N^2$ operations, and a Fast Fourier Transformation (FFT), which requires $\frac{5}{2} \log_2 N - 3 N$ operations when transforming real variables \citep{canuto2006-book}. The vertical transformation must be computed as a matrix multiplication applied to each of the $N_x N_y/2$ wavenumber vectors of length $N_z$, for total computational cost of $N_z^2 N_x N_y$. The horizontal transformation can be computed using an FFT algorithm applied to each depth $N_z$ for a total computational cost of $\frac{5}{2} N_z N_x N_y \log_2 N_x N_y - 3 N_z N_x N_y$. The transformation from wave-vortex space to physical space requires applying the vertical transformation 10 times (7 for the hydrostatic case) and the horizontal transformation 10 times. To finish the pseudospectral multiplication, the results must be projected back onto wave-vortex space for a grand total of 13 horizontal transformations and 10 vertical transformations. Assuming that $N_x = N_y$, the total computational cost of the horizontal and vertical transforms are approximately equal when $10 \log_2 N_x = N_z$, or $13 \log_2 N_x = N_z$ for the hydrostatic case. This means that for a horizontal resolution of $256^2$ the horizontal transformations dominate the computation time until about 80-100 vertical modes are used.


%
\section{Conclusion}
\label{sec:conclusion}
%

The wave-vortex decomposition presented in this paper unambiguously separates linear wave and geostrophic motions under arbitrary stratification into decoupled modes at any given instant in time. The present decomposition has been fully implemented for arbitrary stratification, as well as the special case of constant stratification (see Appendices \ref{sec:constant-stratification} and \ref{sec:numerical-implementation}). The methodology has been validated against output from a linear simulation of a Boussinesq model by confirming that the initial conditions can be exactly recovered at all output times.  We have further shown that this method successfully reproduces the results of more traditional methods that rely on spatial-temporal filtering.  

In addition to these basic validations, the hydrostatic nonlinear equations of motion projected in wave-vortex space (see appendix \ref{sec:nonlinear-eom}) are shown to successfully reproduce changes in wave-vortex amplitude from a nonlinear Boussinesq model.  This suggests that the nonlinear equations of motion can be integrated in wave-vortex space with little modification from the work presented here. Estimates of the computational complexity of the method, presented in section \ref{sec:modeling}, show that numerical integration of the wave-vortex modes may actually perform better than integration in physical space when the stratification profile varies strongly with depth.

One of the more useful aspects of the decomposition is the ability to determine the exact nonlinear pathways that move energy between the wave and geostrophic solutions at different scales. This can be done diagnostically (as in section \ref{sec:applications}) or while directly integrating the equations of motion. By selectively turning off transfer mechanisms, one can also derive reduced-interaction models, such as in \citet{hernandez-duenas2014-jfm}, but now also for vertically bounded flows with variable stratification.

While the present work is a step towards separating wave and vortex motions in generalized flows, there are still limitations to this methodology that prevent its application to, for example, output from global circulation models. In particular, the flows considered here currently lack 1) surface and bottom buoyancy anomalies 2) a free surface, 3) horizontal dependence on stratification or background mean flow, and 4) bottom topography. Two recent studies offer significant progress towards addressing the first two issues. First, \citet{smith2013-jpo} showed that geostrophic motions can be decomposed into uncoupled structures that include a surface buoyancy anomaly, and are also orthogonal (or `diagonalize energy' to use the terminology therein), albeit under rigid lid conditions. Second, \citet{kelly2016-jpo} showed that in the presence of a free surface, there exists a complete set of modes as well as an orthogonality condition that decouples the surface mode from the interior modes, allowing for an unambiguous partitioning of wave energy. Taken together, these results suggest that it should be possible to create a complete framework that includes both a free surface and a surface buoyancy anomaly.\\

We would like to acknowledge National Science Foundation awards OCE-1658564, OCE-1536747, and OCE-1536439 for funding this research. We thank Peter Bartello and 2 anonymous reviewers for helpful suggestions that improved the manuscript. Declaration of Interests: The authors report no conflict of interest.



%
\appendix
\section{Constant stratification}
\label{sec:constant-stratification}
%

Assuming constant stratification, $N^2(z) = N_0^2$, significantly simplifies the problem because the $F$-modes become orthogonal. The vertical modes take the form,
\begin{subequations}
\begin{align} \label{baroclinic_g_mode}
G^{N_0}_j(z) =& A  \sin \left( m_j ( z + D) \right)\\ \label{baroclinic_f_mode}
F^{N_0}_j(z) =& A h_j m_j \cos \left( m_j ( z + D) \right).
\end{align}
\end{subequations}{}
with eigendepth $h_j=\frac{1}{g} \frac{N_0^2 - f_0^2}{k^2+l^2+m_j^2}$ and vertical wavenumber $m_j = \frac{j\pi}{D}$. 
Using the normalization $A^2 = \frac{1}{D}\frac{2g}{N^2 - f^2}$ results in the following orthogonality conditions,
\begin{subequations}
\begin{align}
    \frac{1}{g}\int_{-D}^0 (N_0^2 - f_0^2) G^{N_0}_i(z) G^{N_0}_j(z) \, dz = \delta_{ij} & \\
    \int_{-D}^0 F^{N_0}_i(z) F^{N_0}_j(z) \, dz =  \frac{g h_j^2 m_j^2}{N_0^2 - f_0^2} \delta_{ij}. &
\end{align}{}
\end{subequations}{}

\noindent
The orthogonality conditions imply that if $\eta(z) = \sum_n \hat{\eta}_n G^{N_0}(z)$ or $u(z) = \sum_n \hat{u}_n F^{N_0}(z)$, then
\begin{subequations}
\label{const_strat_transform}
\begin{align}
    \hat{\eta}_n =& \frac{N_0^2 - f_0^2}{g} \int_{-D}^0 G^{N_0}_n(z) \eta(z) \, dz \\
    \hat{u}_n =& \frac{N_0^2 - f_0^2}{g h_n^2 m_n^2} \int_{-D}^0 F^{N_0}_n(z) u(z) \, dz.
\end{align}{}
\end{subequations}{}

\noindent
The consequence is that $(\hat{u},\hat{v})$ can be recovered directly, without integrating to get transport quantities, that is \eqref{wave_coeff_soln}--\eqref{vortex_coeff_soln} with $\tilde{\Delta}(t)$ replaced by $\tilde{\delta}(t)$, $\tilde{\Pi}(t)$ replaced by $\tilde{PV}(t)$, and $\tilde{A_\pm}$ no longer normalized by $h$. 
\begin{align}
\hat{A}_+ =& \frac{e^{-i \omega t}}{2 K} \left[  - i\tilde{\delta}(t) - \frac{1}{\omega} \left( f \tilde{\textrm{PV}}(t) + \frac{\omega^2}{h} \tilde{\eta}(t) \right) \right]  \\
\hat{A}_- =& \frac{e^{i \omega t}}{2 K} \left[  - i\tilde{\delta}(t) + \frac{1}{\omega} \left( f \tilde{\textrm{PV}}(t) + \frac{\omega^2}{h} \tilde{\eta}(t) \right) \right] \\ 
\hat{A}_0 =& - \frac{f_0 h}{\omega^2} \tilde{\textrm{PV}}(t)
\end{align}
where,
\begin{multline}
\tilde{\delta}(t) =   i \left( k \tilde{u}(t) + l \tilde{v}(t) \right), \\  \tilde{PV}(t) \equiv i \left( k \tilde{v}(t) - l \tilde{u}(t) \right) - f_0 \tilde{\eta}(t)/h.
\end{multline}

%
\section{Numerical implementation}
\label{sec:numerical-implementation}
%

The decomposition was tested with output from a linear simulation with a rotating spectral Boussinesq model \citep{winters2012-om} with constant stratification. Implementation of the methodology requires the following steps,
\begin{enumerate}
    \item Discrete Fourier transforms of $u$, $v$, and $\eta$ in $x$ and $y$.
    \item A discrete cosine transform of $u$ and $v$ and a discrete sine transform of $\eta$ in $z$.
    \item Computation of the wave-vortex coefficients from the transformed variables.
\end{enumerate}{}
The last step requires careful bookkeeping to ensure that all terms are properly accounted for and not double-counted.

The domain is assumed to have $[N_x\times N_y \times (N_z+1)]$ points, where $N_x$ and $N_y$ are typically in powers of 2 to take advantage of the fast Fourier transform (FFT), while $N_z+1$ has $2^{n}+1$ points to accommodate the type-I discrete cosine transforms (DCT-I) and type-I discrete sine transforms (DST-I) used by the Winters model.

\subsection{Horizontal transformation}
\label{horizontal_transform}

The finite-length Fourier transform in a periodic domain is given by
\begin{equation}
\mathcal{F}\left[ f(x) \right] = \frac{1}{L} \int_{-D}^0 f(x) e^{-i k_j x} \, dx
\end{equation}
where $k_j=\frac{2\pi j}{L}$. For a discretized domain with points at $x_n=n \Delta$ where $n=[0 \,...\, N-1]$ and $\Delta=L/N$, the discrete Fourier transform is
\begin{equation}
\hat{f}(k_j) = \mathcal{DFT}\left[ f(x_n) \right] = \frac{1}{L} \sum_{n=0}^{N-1} f(x_n) e^{-i k_j x_n} \Delta.
\end{equation}
with wavenumbers $k_j=\frac{2\pi j}{L}$ now limited to $j=[0 \,...\, N-1]$. Variance is preserved following Plancherel's theorem,
\begin{equation}
    \frac{1}{N}\sum_{n=0}^{N-1} |f(x_n)|^2 = \sum_{j=0}^{N-1} S(k_j) dk
\end{equation}{}
where $dk=\frac{2\pi}{L}$ and $S(k_j)=\frac{L}{2\pi} |\hat{f}(k_j)|^2$ is defined as the spectrum.

\newcolumntype{g}{>{\columncolor{LightGray}}c}
\begin{table}
\begin{center}\begin{tabular}{|c|c|c|c|c|g|g|g|}\hline 
 \cyan 0,0 & 0,1 & 0,2 & 0,3 & \orange 0,-4 & 0,-3 & 0,-3 & 0,-1\\\hline
 1,0 & 1,1 & 1,2 & 1,3 & \orange 1,-4 & 1,-3 & 1,-3 & 1,-1\\\hline
 2,0 & 2,1 & 2,2 & 2,3 &  \orange 2,-4 & 2,-3 & 2,-3 & 2,-1\\\hline
 3,0 & 3,1 & 3,2 & 3,3 &  \orange 3,-4 & 3,-3 & 3,-3 & 3,-1 \\\hline
\orange -4,0 & \orange -4,1 & \orange -4,2 & \orange -4,3 & \orange  -4,-4 & -4,-3 & -4,-3 & -4,-1 \\\hline 
\pink -3,0 & -3,1 & -3,2 & -3,3 &  \orange -3,-4 & -3,-3 & -3,-3 & -3,-1\\\hline 
\pink -2,0 & -2,1 & -2,2 & -2,3 &  \orange -2,-4 & -2,-3 & -2,-3 & -2,-1\\\hline 
\pink -1,0 & -1,1 & -1,2 & -1,3 &  \orange -1,-4 & -1,-3 & -1,-3 & -1,-1\\\hline
         \end{tabular}
         \caption{Table of FFT coefficients for wavenumbers (k,l). We have let $-4 \leq k \leq 3$ and $-4 \leq l \leq 3$, consistent with an 8x8 2D FFT. The grey shaded region shows the redundant coefficients that are determined from Hermitian conjugacy by changing the sign on the $l$ component. The pink shaded components are Hermitian conjugate by changing the sign on the $k$ component. The orange components are Nyquist components, and thus not full resolved. The cyan shaded component (including (k,l)=(-4,0), (-4,4) and (0,-4)) are self-symmetric, and therefore strictly real}
\end{center}
\label{fft-table}
\end{table}

Applying the $\mathcal{DFT}$ in both $x$ and $y$ using the usual numerical algorithms on a real value function results in a two-dimensional transformed matrix as shown in table \ref{fft-table}. For a real valued function the power is split between the two conjugate pairs, and therefore $\hat{f}(k_j)$ has to be doubled to be compared to $\tilde{f}_{jkl}$ in \eqref{f_eqn_form}. The grey and pink regions in table \ref{fft-table} are Hermitian conjugates of other values in the table. The Nyquist frequency $j=N/2$ is unresolved since the sine at the Nyquist is zero, and thus the orange regions are also ignored. Only the white regions and the cyan component at $k=l=0$ contain the information for the inversion.

\subsection{Vertical transformation}
The finite-length sine and cosine transforms are given by
\begin{equation}
\label{st}
\mathcal{S}\left[ g(z) \right] = \frac{2}{D} \int_{-D}^0 g(z) \sin( m_j z) \, dz
\end{equation}
and 
\begin{equation}
\label{ct}
\mathcal{C}\left[ f(z) \right] = \frac{2}{D} \int_{-D}^0 f(z) \cos( m_j z) \, dz
\end{equation}
where $m_j = \frac{j \pi}{D}$. The discretized versions of these transforms, the DST-I and DCT-I used by the Winters model, are defined with points at $z_n=n\Delta$ where $n=0..N_z$ and $\Delta=D/N_z$. Note that this differs from the discretization for the $\mathcal{DFT}$ by including endpoints. This choice of discretization results in the discrete transforms
\begin{equation}
\hat{g}(m_j) = \mathcal{DST}\left[ g(z_n) \right] = \frac{2}{D} \sum_{n=0}^{N_z} g(z_n) \sin( m_j z_n ) \Delta
\end{equation}
and
\begin{multline}
\label{dct}
\hat{f}(m_j) = \mathcal{DCT}\left[ f(z_n) \right] =  \frac{2}{D}\Bigg(  \frac{f(0)}{2} + \\  \sum_{n=1}^{N_z-1} f(z_n) \cos( m_j z_n )  + (-1)^{j} \frac{f(D)}{2}\Bigg) \Delta.
\end{multline}
with vertical wavenumbers at $m_j = \frac{j \pi}{D}$ where $j=0..N_z$.  The sum in the $\mathcal{DCT}$ treats the end points separately, as they have only half the width of the other points, $\Delta/2$. For the $\mathcal{DST}$ the function is zero at the endpoints, $g(z_0)=g(z_n)=0$, and the $m_0$ and $m_{N_z}$ wavenumber components have zero power.

With these definitions of the transform, Plancherel's Theorem states that,
\begin{equation}
    \frac{1}{N_z}\sum_{n=1}^{N_z-1} |g(x_n)|^2 = \sum_{j=1}^{N_z-1} S(m_j) dm
\end{equation}{}
with $S(m_j)=\frac{D}{2\pi} |\hat{g}(m_j)|^2$ and
\begin{multline}
    \frac{1}{N_z} \left( \frac{|f(0)|^2}{2} +  \sum_{n=1}^{N_z-1} |f(z_n)|^2 +  \frac{|f(D)|^2}{2} \right) \\= \left(\frac{S(m_0)}{2} + \sum_{j=1}^{N_z-1} S(m_j) + 2S(m_{Nz}) \right) dm
\end{multline}{}
with $S(m_j)=\frac{D}{2\pi} |\hat{f}(m_j)|^2$ where $dm=\frac{\pi}{D}$. The variance of the $m_0$ wavenumber is notably a factor of 2 larger than the variance of a constant function.

\subsection{Wave-vortex coefficients}

Applying the discrete transformations exactly as defined above to $u$, $v$ and $\eta$ results in the following matrices,
\begin{subequations}
\begin{align}
    \hat{u}_{klj} =&  \mathcal{DCT}_z \left[\mathcal{DFT}_y \left[\mathcal{DFT}_x \left[ u(x,y,z) \right] \right] \right] \\
    \hat{v}_{klj} =& \mathcal{DCT}_z \left[\mathcal{DFT}_y \left[\mathcal{DFT}_x \left[ v(x,y,z) \right] \right] \right] \\
    \hat{\eta}_{klj} =& \mathcal{DST}_z \left[\mathcal{DFT}_y \left[\mathcal{DFT}_x \left[ \eta(x,y,z) \right] \right] \right].
\end{align}{}
\end{subequations}

\subsubsection{Coefficients, $k^2 + l^2 > 0$, $j=0$}

Starting with the $j=0$ mode, the discrete transforms give nonzero coefficient functions for $\hat{u}_{kl0}$ and $\hat{v}_{kl0}$, but zero for $\hat{\eta}_{kl0}$. The $\mathcal{DCT}$ inflates the power by a factor of two, but the $\mathcal{DFT}$ returns only half the power of the real-value function. The result is that,
\begin{equation}
    \hat{A}_0 = - i \frac{f_0}{g K^2} \left( k \hat{v}_{kl0} - l \hat{u}_{kl0} \right)
\end{equation}{}
exactly as written before.

\subsubsection{Coefficients, $k^2 + l^2 > 0$, $j>0$}

The projection operations as defined in \eqref{const_strat_transform} can be related to the sine and cosine transformations in \eqref{st} and \eqref{ct} by a scaling,
\begin{subequations}
\begin{align}
    \bar{u}_{klj} =& \frac{\sqrt{D(N_0^2 - f_0^2)}}{h_{klj} m_{j} \sqrt{2g} } \hat{u}_{klj} \\
    \bar{v}_{klj} =& \frac{\sqrt{D(N_0^2 - f_0^2)}}{h_{klj} m_{j}  \sqrt{2g}} \hat{v}_{klj} \\
    \bar{\eta}_{klj} =& \frac{\sqrt{D(N_0^2 - f_0^2)}}{\sqrt{2g}} \hat{\eta}_{klj}.
\end{align}{}
\end{subequations}
The wave-vortex coefficients are then recovered with,
\begin{align}
    \hat{A}_\pm =& \frac{e^{\mp i\omega t}}{2} \left[ \frac{k \omega \pm i l f_0}{\omega K} \bar{u}_{klj} + \frac{l \omega \mp i k f_0}{\omega K}\bar{v}_{klj} \pm \frac{gK}{\omega} \bar{\eta}_{klj} \right] \\
    \hat{A}_0 =& i \frac{l h f_0}{\omega^2} \bar{u}_{klj} - i \frac{k h f_0}{\omega^2} \bar{v}_{klj} + \frac{ f_0^2}{ \omega^2} \bar{\eta}_{klj}
\end{align}{}.

%
\section{Nonlinear equations of motion}
\label{sec:nonlinear-eom}
%
Here we rewrite the hydrostatic nonlinear equations of motion for $(u,v,\eta,w,p)$ as nonlinear equations of motion for $(A_+,A_-,A_0)$.

\subsection{Linear transforms}

We separate the linear transformations into two parts $S \cdot T_\omega$ where $S : (A_+,A_-,A_0) \to (u,v,\eta,w,p)$ maps from wave-vortex space to physical variables and $T_\omega : (A_+,A_-,A_0) \to (A_+,A_-,A_0)$ winds the wave phases from the initial conditions to the current time in wave-vortex space. The total transformations are therefore defined so that
\begin{multline}
\label{linear-transformations}
\left[\begin{array}{c}
u \\
v \\
\eta \\
w \\
p
\end{array} \right] = S \cdot T_\omega
\left[\begin{array}{c} \hat{A}_+  \\  \hat{A}_-  \\\hat{A}_0 \end{array}\right]
\textrm{ and } \\
\left[\begin{array}{c} \hat{A}_+  \\  \hat{A}_-  \\\hat{A}_0 \end{array}\right]
 = T_\omega^{-1} S^{-1}
\left[\begin{array}{c}u \\v \\ \eta \end{array}\right],
\end{multline}
where it is understood that the physical variables are functions of $(x,y,z,t)$ and wave-vortex coefficients are functions of $t$ and indexed with $jkl$. The operator $T_\omega$ is the phase winding operator,
\begin{equation}
    T_\omega = 
    \left[\begin{array}{ccc}
    e^{i\omega t} & 0 & 0 \\
    0 & e^{-i\omega t} & 0 \\
    0 & 0 & 1
    \end{array}\right]
\end{equation}
with inverse $T_\omega^{-1}$ given by,
\begin{equation}
    T_\omega^{-1} = 
    \left[\begin{array}{ccc}
    e^{-i\omega t} & 0 & 0 \\
    0 & e^{i\omega t} & 0 \\
    0 & 0 & 1
    \end{array}\right].
\end{equation}
The projection onto physical variables is defined by,
\begin{multline}
S = 
    \mathcal{DFT}^{-1} \cdot 
    \left[\begin{array}{ccccc}
    \mathcal{F}^{-1} & 0 & 0 & 0 & 0 \\
    0 & \mathcal{F}^{-1}  & 0 & 0 & 0 \\
    0 & 0 & \mathcal{G}^{-1}  & 0 & 0 \\
    0 & 0 & 0 & \mathcal{G}^{-1}  & 0 \\
    0 & 0 & 0 & 0 & \mathcal{F}^{-1}
    \end{array}\right] \\
\cdot \left[\begin{array}{ccc} \frac{k\omega - il f_0}{\omega K} & \frac{k\omega + il f_0}{\omega K} & -i \frac{g}{f_0} l \\
\frac{l \omega + i k f_0}{\omega K } & \frac{l\omega - i k f_0}{\omega K } & i\frac{g}{f_0} k  \\
- \frac{K h}{\omega} &  \frac{K h}{\omega} & 1 \\
-iKh & -iKh & 0 \\
-\rho_0 g \frac{K h}{\omega} & \rho_0 g \frac{K h}{\omega} & \rho_0 g
\end{array}\right]
\end{multline}
with inverse,
\begin{multline}
    S^{-1} = \left[\begin{array}{ccc}
 \frac{k \omega + i l f_0}{2\omega K} & \frac{l \omega - i k f_0}{2\omega K} & - \frac{gK}{2\omega} \\
  \frac{k \omega - i l f_0}{2\omega K} & \frac{l \omega + i k f_0}{2\omega K} &  \frac{gK}{2\omega} \\
  i \frac{l h f_0}{\omega^2}& - i \frac{k h f_0}{\omega^2} &  \frac{ f_0^2}{ \omega^2}
\end{array}\right] \\  \cdot
 \left[\begin{array}{ccc}
 \mathcal{F} & 0 & 0 \\
 0 & \mathcal{F} & 0 \\
 0 & 0 & \mathcal{G}
 \end{array}\right] 
\cdot \mathcal{DFT}.
\end{multline}
These transformations use hydrostatic versions of equations \eqref{wavevortex-forward-transformation} and \eqref{wavevortex-inverse-transformation}, as well as the  $\mathcal{DFT}$as defined in appendix \ref{sec:numerical-implementation}. The vertical projection operators are defined with the hydrostatic modes,
\begin{equation}
\begin{split}
    \mathcal{F} \left[ f(z) \right] \equiv \frac{1}{h_j} \int_{-D}^0  f(z) F^\textrm{HS}_j \, dz, \\
    \mathcal{G} \left[ g(z) \right] \equiv \frac{1}{g} \int_{-D}^0 g(z) N^2(z)  G^\textrm{HS}_j \, dz
    \end{split}
\end{equation}
with inverses
\begin{equation}
\begin{split}
        \mathcal{F}^{-1} \left[ \tilde{f}_{j} \right] \equiv  \sum_{j} \tilde{f}_{j}  F^\textrm{HS}_{j}(z), \\
    \mathcal{G}^{-1} \left[ \tilde{g}_{j} \right] \equiv  \sum_{j} \tilde{g}_{j} G^\textrm{HS}_{j}(z).
        \end{split}
\end{equation}
Note that, unlike the non-hydrostatic case, the vertical projection and $\mathcal{DFT}$ operator commute.

\subsection{Nonlinear equations of motion}

As a preliminary step before re-writing the nonlinear equations of motion, the density equation,
\begin{equation}
     \partial_t \rho + w \partial_z \bar{\rho} + u \partial_x \rho + v \partial_y \rho + w  \partial_z \rho = 0
\end{equation}
can be expressed using $\eta=-\rho/\partial_z\bar{\rho}$ and $N^2=-\frac{g}{\rho_0}\partial_z\bar{\rho}$ as
\begin{equation}
\partial_t \eta - w + u \partial_x\eta + v \partial_y\eta + w \left(\partial_z\eta +\eta \partial_z \ln N^2 \right)  = 0.
\end{equation}

The nonlinear equations of motion written in matrix form are
\begin{equation}
\label{nonlinear-eom}
\begin{split}
\partial_t 
 \underbrace{
\left[\begin{array}{c}
u \\
v \\
\eta \\
0 \\
0
\end{array} \right]}_{\psi}
+
 \underbrace{
    \left[\begin{array}{ccccc}
    0 & -f_0 & 0 & 0 & \frac{1}{\rho_0}\partial_x \\
    f_0 & 0 & 0 & 0 & \frac{1}{\rho_0}\partial_y \\
        0 & 0 & 0 & -1 & 0 \\
    0 & 0 & 1 & 0 & N^{-2}\frac{1}{\rho_0}\partial_z \\
     \partial_x & \partial_y & 0 & \partial_z & 0
    \end{array}\right] }_{\Lambda}
    \underbrace{
    \left[\begin{array}{c}
u \\
v \\
\eta \\
w \\
p
\end{array} \right]}_{\psi} \\+
\underbrace{
\left[\begin{array}{c}
\textrm{uNL} \\
\textrm{vNL} \\
\textrm{nNL} \\
0 \\
0
\end{array} \right]}_{\Lambda^{\textrm{NL}}}\underbrace{
    \left[\begin{array}{c}
u \\
v \\
\eta \\
w \\
p
\end{array} \right]}_{\psi}= 0
\end{split}
\end{equation}{}
where the nonlinear operator $\Lambda^{\textrm{NL}}$ is defined as
\begin{small}
\begin{subequations}
\label{nonlinear-terms}
\begin{align}
    \textrm{uNL}\left[u,v,\eta,w \right]=& u \partial_x u + v \partial_y u + w \partial_z u \\
    \textrm{vNL}\left[u,v,\eta,w \right]=&u \partial_x v + v \partial_y v + w \partial_z v \\
    \textrm{nNL}\left[u,v,\eta,w \right]=& u \partial_x \eta + v \partial_y \eta + w \left(\partial_z \eta +\eta \partial_z \ln N^2 \right)
\end{align}
\end{subequations}
\end{small}
Changing to wave-vortex space, we apply the transformation $T_\omega^{-1} S^{-1}$ to \eqref{nonlinear-eom},
\begin{small}
\begin{align}
    \partial_t \psi + \Lambda \psi + \Lambda^{\textrm{NL}} \psi =& 0 \\
    T_\omega^{-1} S^{-1}\partial_t \psi + T_\omega^{-1} S^{-1}\Lambda \psi + T_\omega^{-1} S^{-1}\Lambda^{\textrm{NL}} \psi =& 0 \\
    \left(T_\omega^{-1} S^{-1}\partial_t S T_\omega\right) \mathbf{A} + \left(T_\omega^{-1} S^{-1}\Lambda S T_\omega\right) \mathbf{A} + T_\omega^{-1} S^{-1}\Lambda^{\textrm{NL}} \psi =& 0
\end{align}
\end{small}
where $\mathbf{A}=T_\omega^{-1} S^{-1} \psi$ as in \eqref{linear-transformations} above. The result of this transformation are the nonlinear equations of motion in wave-vortex space,
\begin{multline}
\label{wave-vortex-nonlinear-eom}
    \partial_t \left[\begin{array}{c} \hat{A}_+  \\  \hat{A}_-  \\\hat{A}_0 \end{array}\right] = \left[\begin{array}{c}
 F_+ \\
 F_- \\
F_0
\end{array} \right]  \textrm{ where } \\
\left[\begin{array}{c}
F_+ \\
F_- \\
F_0
\end{array} \right] \equiv 
    -T_\omega^{-1} S^{-1}\left[\begin{array}{c}
\textrm{uNL} \\
\textrm{vNL} \\
\textrm{nNL} 
\end{array} \right] \psi
\end{multline}
If we had not included the time-winding operator $T_\omega$ in the basis transformation, the equations for the two wave coefficients, $A_\pm$, would include linear terms to evolve the phases, and thus the nonlinear equations would resemble forced wave equations. It is also worth noting that the first part of this transformation, the projection onto vertical modes, is identical to \citet{kelly2016-jpo}, although without the free-surface.

The physical variables $(u,v,\eta,w)$ in \eqref{wave-vortex-nonlinear-eom} can be expressed in terms of wave-vortex coefficients $(A_+,A_-,A_0)$, in which case the multiplication of physical variables becomes a convolution of wave-vortex coefficients. However, we have instead left this in pseudospectral form, where the multiplication occurs in physical space and the result is transformed into wave-vortex space.

\subsection{Energy flux}

The nonlinear equations of motion \eqref{wave-vortex-nonlinear-eom} each have the form $\partial_t A =  F$, where $F$ is a complicated function of the other dependent variables. To compute the flux diagnostically, we take an Euler time step, for which the solution is $A(t+\Delta t) = F(t)\Delta t + A(t)$. The change in energy is computed by multiplying by the complex conjugate and taking the time derivative so that,
\begin{equation}
    \frac{d}{dt} A^2 = 2 F \bar{F} \Delta t + 2 \mathcal{R} \left( F \bar{A} \right).
\end{equation}
The first term, dependent on the Euler time step $\Delta t$, is only of consequence when $A$ is zero and can be neglected for these diagnostics. The squared amplitude is converted into total energy using the expression in Section \ref{sec:solutions}, resulting in expressions for the change in total wave and geostrophic energies,
\begin{equation}
\begin{split}
\label{energy-flux}
    \frac{d}{dt} E_\pm = \frac{1}{2} h \mathcal{R} \left( F_\pm \bar{A}_\pm \right) \textrm{ and } \\\frac{d}{dt} E_0  = \frac{1}{2} g\left( 1+ \frac{ghK^2}{f_0^2}\right) \mathcal{R} \left( F_0 \bar{A}_0 \right).
    \end{split}
\end{equation}

\subsection{Nonlinear pathways}

While the total energy flux is computed using \eqref{energy-flux} with \eqref{wave-vortex-nonlinear-eom}, this does not immediately tell us \emph{which} nonlinear interactions are important. Writing out the non-linear terms from \eqref{wave-vortex-nonlinear-eom} more explicitly, we have
\begin{strip}
\begin{subequations}
\label{nonlinear-rhs}
\begin{align}
\label{wave_p_forcing}
    e^{i\omega t}F_+ =& -\frac{1}{2 K} \left( k \cdot \widehat{\textrm{uNL}} + l \cdot \widehat{\textrm{vNL}} \right) - i\frac{f_0}{2\omega K} \left( l \cdot \widehat{\textrm{uNL}} - k \cdot \widehat{\textrm{vNL}} \right) + \frac{g K}{2 \omega}\widehat{\textrm{nNL}} \\ \label{wave_m_forcing}
    e^{-i\omega t}F_- =& -\frac{1}{2 K} \left( k \cdot \widehat{\textrm{uNL}} + l \cdot \widehat{\textrm{vNL}} \right) + i\frac{f_0}{2\omega K} \left( l \cdot \widehat{\textrm{uNL}} - k \cdot \widehat{\textrm{vNL}} \right) - \frac{g K}{2 \omega}\widehat{\textrm{nNL}}\\
    F_0 =& -i h \frac{  f_0}{\omega^2} \left( l \cdot \widehat{\textrm{uNL}} - k \cdot \widehat{\textrm{vNL}} \right) - \frac{f_0^2}{ \omega^2}\widehat{\textrm{nNL}}
\end{align}
\end{subequations}
\end{strip}
where we have simplified the notation so that $\widehat{\textrm{uNL}}=\mathcal{F} \cdot \mathcal{DFT}[\textrm{uNL}]$ and $\widehat{\textrm{nNL}}=\mathcal{G} \cdot \mathcal{DFT}[\textrm{nNL}]$. The idea now is to single out the specific nonlinear pathways that are move energy between modes.

Using the wave-vortex decomposition, the physical variables in uNL, vNL, and nNL can be expressed in terms of their constituent parts. In this case we separate the inertial oscillations (io) from gravity waves (igw), and combine the the positive and negative waves into a single internal gravity wave part. With the addition of the geostrophic component (g), the physical variables are unambigiously separated as,
\begin{subequations}
\begin{align}
    u =& u_\textrm{io} + u_\textrm{igw} + u_\textrm{g} \\
    v =& v_\textrm{io} + v_\textrm{igw} + v_\textrm{g} \\
    \eta =& \eta_\textrm{igw} + \eta_\textrm{g} \\
    w =& w_\textrm{igw}.
\end{align}
\end{subequations}

The primary hypothesis described in Section \ref{sec:applications} is that the advection of geostrophic vorticity by inertial oscillations is the energy source for internal gravity waves. This particular nonlinear pathway is part of the second term in parenthesis in \eqref{wave_p_forcing} and \eqref{wave_m_forcing},
\begin{multline}
    F_\pm^{\textrm{io}\nabla\textrm{g}} = \pm \frac{i f_0  e^{\mp i\omega t}}{2 \omega K} \big( k \cdot \mathcal{F} \cdot \mathcal{DFT}\left[u_\textrm{io} \partial_x v_\textrm{g} + v_\textrm{io} \partial_y v_\textrm{g} \right] \\- l \cdot \mathcal{F} \cdot \mathcal{DFT}\left[u_\textrm{io} \partial_x u_\textrm{g} + v_\textrm{io} \partial_y u_\textrm{g} \right] \big),
\end{multline}
and can be further simplified to
\begin{small}
\begin{align}
    F_\pm^{\textrm{io}\nabla\textrm{g}} =& \pm \frac{f_0  e^{\mp i\omega t}}{2 \omega K} \left( \mathcal{F} \cdot \mathcal{DFT}\left[u_\textrm{io} \partial_x \zeta_\textrm{g} + v_\textrm{io} \partial_y \zeta_\textrm{g} \right] \right),
\end{align}
\end{small}
where $\zeta_g \equiv \partial_x v_g - \partial_y u_g$.

It is useful to treat the forcing of inertial oscillations separately from the rest of the waves because so many terms cancel. The forcing term can be found by considering the negative wave forcing \eqref{wave_m_forcing}, or with the transformation,
\begin{equation}
S^{-1} = \frac{1}{2}
    \left[\begin{array}{ccc}
    0 & 0 & 0 \\
    1 & i & 0 \\
    0 & 0 & 0 \\
    \end{array}\right].
\end{equation}
The total forcing on inertial waves is thus
\begin{equation}
     F_\textrm{io} = -\frac{1}{2} e^{if_0 t} \left( \widehat{\textrm{uNL}} + i  \widehat{\textrm{vNL}} \right).
\end{equation}
Individual nonlinear pathways can be computed by considering the constituent parts. The two most relevant pathways for this problem are
\begin{equation}
     F_\textrm{io}^{\textrm{igw}\nabla\textrm{g}} = -\frac{1}{2} e^{if_0 t} \mathcal{F} \left[ \overline{ \mathbf{u}_\textrm{igw} \cdot \nabla \left( u_\textrm{g} + i v_\textrm{g} \right) } \right]
\end{equation}
and
\begin{equation}
     F_\textrm{io}^{\textrm{igw}\nabla\textrm{igw}} = -\frac{1}{2} e^{if_0 t} \mathcal{F} \left[ \overline{ \mathbf{u}_\textrm{igw} \cdot \nabla \left( u_\textrm{igw} + i v_\textrm{igw} \right) } \right],
\end{equation}
where the $\overline{(\cdot)}$ indicates horizontal averaging, equivalent to the $K=0$ component of the $\mathcal{DFT}$ operator that it replaces.

\bibliographystyle{ametsoc2014}

\end{document}